\documentclass[10 pt,twocolumn, twoside, journal]{IEEEtran}

\ifCLASSINFOpdf

\else

\fi
\usepackage{graphicx}
\usepackage{subfig}
\usepackage{amsmath}
\usepackage{amssymb}
\usepackage{titlesec}
\usepackage{cite}
\usepackage{xcolor}
\usepackage{paralist}
\usepackage[utf8]{inputenc}
\usepackage[english]{babel}
\usepackage{amsthm}

\usepackage[shortlabels]{enumitem}
\usepackage{url}
\graphicspath{{figures/}}
\begin{document}
%
\title{Injecting Reliable Radio Frequency Fingerprints Using Metasurface for The Internet of Things}

\author{Sekhar~Rajendran, \IEEEmembership{Student Member, IEEE,}
Zhi~Sun, \IEEEmembership{Senior Member, IEEE,}
Feng~Lin, \IEEEmembership{Senior Member, IEEE,}
and Kui~Ren \IEEEmembership{Fellow, IEEE}

\thanks{S.~Rajendran and Z.~Sun are with the Department
of Electrical Engineering, University at Buffalo, Buffalo, NY, 14226 USA, e-mail: sekharra@buffalo.edu, zhisun@buffalo.edu}
\thanks{F.~Lin and K.~Ren are with the Institute of Cyberspace Research, Zhejiang University, Hangzhou, Zhejiang, 310058 China, e-mail: flin@zju.edu.cn, kuiren@zju.edu.cn}}

\maketitle

\begin{abstract}
In Internet of Things, where billions of devices with limited resources are communicating with each other, security has become a major stumbling block affecting the progress of this technology. Existing authentication schemes based on digital signatures have overhead costs associated with them in terms of computation time, battery power, bandwidth, memory, and related hardware costs. Radio frequency fingerprint (RFF), utilizing the unique device-based information, can be a promising solution for IoT. However, traditional RFFs have become obsolete because of low reliability and reduced user capability. Our proposed solution, Metasurface RF-Fingerprinting Injection (MeRFFI), is to inject a carefully-designed radio frequency fingerprint into the wireless physical layer that can increase the security of a stationary IoT device with minimal overhead. The injection of fingerprint is implemented using a low cost metasurface developed and fabricated in our lab, which is designed to make small but detectable perturbations in the specific frequency band in which the IoT devices are communicating. We have conducted comprehensive system evaluations including distance, orientation, multiple channels where the feasibility, effectiveness, and reliability of these fingerprints are validated. The proposed MeRFFI system can be easily integrated into the existing authentication schemes. The security vulnerabilities are analyzed for some of the most threatening wireless physical layer-based attacks.
\end{abstract}


%
\IEEEpeerreviewmaketitle

\section{Introduction}
Internet of things (IoT) is revolutionizing our daily life and gradually overwhelming over existing IT devices, such as PCs and mobiles, by their ubiquitous connectivity and diverse applications. It is forecasted that by 2022 around 18 billion connected devices will be related to IoT \cite{iot}. The global market for IoT technology has reached \$190 billion in 2018, and it is projected to reach \$1012 billion by 2026 according to a data and analytics company \cite{globaldata}. However, most IoT devices are operating without a secure authentication procedure which makes them vulnerable to cyber attacks.



\begin{figure}[t]
	\centering 
	\includegraphics[scale=0.15]{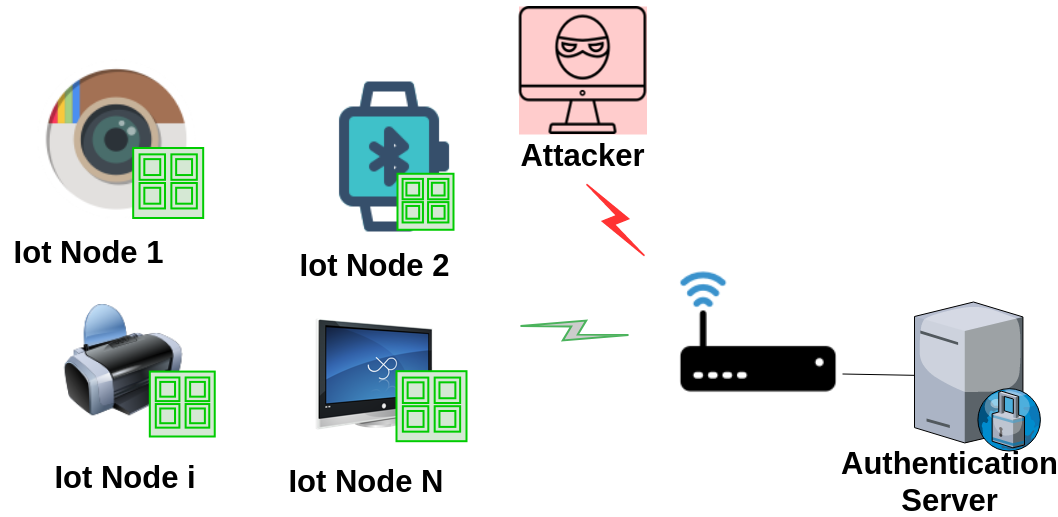}
	\caption{IoT Network in which all devices are equipped with MeRFFI, and an adversary trying to access the network.}
	\label{fig:simplepic}
	\vspace{-15pt}
\end{figure}
The  current  state-of-the-art  IoT  connectivity  technologies employ  cryptographic  protocols \cite{Singh2017} which are vulnerable to attacks \cite{White2014}. Increasing computational complexity of the algorithms is an obvious but costly solution. A promising cost-effective solution to tackle the aforementioned challenge for IoT came in the form of radio frequency fingerprinting (RFF) \cite{Fran19}, which can be observed from the electromagnetic waves radiated by the transmitter. By verifying the RF-fingerprint, secure authentication can be established between IoT nodes. 
RFF is a type of device hardware-based fingerprinting, which is built on the fact that the fabrication process of the transmitter, especially, analog components, introduces certain distinctive imperfections that can act as the fingerprint of the device. To be qualified as a fingerprint, the RFF has to be unique, reliable, and unforgeable to be adopted into practical IoT systems. However, traditional RFF methods are becoming less prominent due to advanced manufacturing techniques, which results in certain limitations: a) \textbf{lack of configurability}-- users have no control over the generation of fingerprint; b) \textbf{limited distinguishability}-- extracted RFFs from different devices may exhibit similar properties with each other because of the reduced user capacity; c) \textbf{low reliability}-- the RFFS are  minuscule and when transmitted over the channel gets diminished, distorted and corrupted by noise d) \textbf{complicated data processing}-- more computation is needed to mine inconspicuous device features caused by the reduced feature space.

To address the issues of low reliability, reduced user capacity, diminishing distinguishability,  and complications in data processing of RFFs,  we propose a novel injectable RFF scheme through electromagnetic metasurfaces (also termed Reconfigurable Intelligent Surfaces \cite{Basar19} or Intelligent Reflective Surfaces \cite{Ozdo20}). Specifically, we designed and developed a system called metasurface for radiofrequency fingerprint injection (MeRFFI). MeRFFI is a metasurface  \cite{Bayatpur2009} designed to make small but detectable perturbations in the specific frequency band in which the IoT devices are communicating. The perturbations created by MeRFFI are multiple well-intended constructive interferences on the transmitter's emission, which do not impact the wireless communication. MeRFFI is intended to be used for stationary IoT devices that use wideband RF channels for communication, making it useful for systems that require high security like printers, health gadgets, wireless cameras, backhaul of sensor networks, and industrial monitoring systems. MeRFFI is low cost (less than 20 cents) and consumes very little power (less than 50 millijoules per code), and its form factor is small enough to be conveniently attached to the IoT transmitters (e.g., as part of transmitter casing, cellphone cover or a health watch's backplane). In this paper, we introduce MeRFFI specific to static application scenarios and discuss the possibility of further broadening its scope to everyday mobile applications like health watches and cell phones.

We present our system with two modes of operation, the passive (fixed pattern) mode provides easy and low-cost access to IoT devices that requires no further processing, while the active (adjustable pattern) mode can provide a time-varying signature by programmatically changing the electromagnetic properties of the metasurface with a microcontroller unit. 
Our key contributions are:
\begin{compactitem}
 \item To the best of our knowledge, we are the first to propose and design a metasurface-based RFF injection system, which overcomes the problems of low reliability and reduced user capacity from traditional RFFs. We have modeled the metasurface enhanced signature injection scheme and further provided potential attack scenarios and security analysis. 
\item We developed and implemented a low-cost MeRFFI prototype. With comprehensive evaluations, including distance, orientation, and multiple channels on the system, we demonstrate the feasibility of MeRFFI as a realistic security signature for IoT authentication.
\item The developed MeRFFI has an intrinsic flexible property that can be utilized in different applications based on different levels of security, cost restraints, and processing capabilities. We demonstrated how MeRFFI could be used in an IoT scenario using three easily integrable authentication protocols.
\end{compactitem}

\section{Related Work} \label{related}
In this section,  two predominantly used  schemes for IoT authentication are discussed \textit{Radio Frequency fingerprinting} and \textit{hardware signature embedding}.
\vspace{-7pt}
\subsection{RF-Fingerprinting (RFF): } The cost constraints on the manufacturing process of IoT transmitters have made the intrinsic variations caused by the components in it, unavoidable. In a typical Direct Conversion transmitter \cite{choi2014}, which is commonly used in IoT transmitters, RFFs are produced by most of its functional components.

 The digital to analog converters are often affected by \textit{integral non-linearity (INL)}. These nonlinear effects have been used as fingerprints \cite{Polak2011}. Several variations can arise from the use of different signal processing techniques \cite{Brik2008} in the transmitting hardware. These cannot be eliminated from the manufacturing process as they depend upon variable tolerances of the different device components. In the mixer part, the in-phase and the quadrature components are required to be at a specific phase angle apart (e.g., 90\textdegree). This \textit{imbalance between the in-phase and quadrature components} have also been leveraged for RF fingerprinting \cite{Zhuo2017}. The \textit{nonlinear characteristics of a PA} often vary widely according to its average output \cite{kwon2010}, as the heat dissipation would alter the chip temperature and hence there would be device dependent signatures \cite{Polak2015} present in the electromagnetic waves emitted by the transmitter. The RF front end that is the drive circuit along with the antenna and its effects on polarization and voltage has also been studied for RFFs \cite{danevrfid2009}. \textit{Clock Jitter} is another RFF  \cite{zanetti2010}, which remains fairly consistent over time, but the clock skews vary significantly across devices. Apart from these, observations have been made on \textit{turn-on and turn-off transient} of the communication device. The energy envelope of the instantaneous transient signal has been researched as RFFs \cite{zanetti2010}.
 
 These RFFs have known vulnerabilities to two main types of replay attacks \cite{Danev2010}, the signal replay attack (receive and replay without modification), and feature replay attack (receive, analyze, generate and transmit).  When the RF-fingerprinting features, modulation based RFF, and transient-based RFF  were tested with feature replay attacks, it was found that modulation based RFFs can be feature replayed with almost 100 percent accuracy. 
This happens because of the \textit{simplicity} of the modulation based features.
Transient-based features
have proven to be more effective. However,  commercial wireless devices cannot capture the transient-based features as they do not have the required dynamic operating frequency range (They are generally processed from high frequencies). \vspace{-7pt}
 \subsection{Hardware Signature Embedding:} This method  uses intrinsic defect based properties to generate true random and unique numbers and apply these as a replacement for cryptographic codes. The most popular solution that came using this principle, applies physically unclonable functions (PUFs). PUF maps input challenges to output responses based on a function determined by inherent variations of that circuit. These responses are used in place of digital signature and are included in the communication packet transmitted by IoT devices for securing the communication between them \cite{urbi2018}. The PUF based solutions are not really physical layer fingerprinting as they are digitally transmitting the signature with the sent packet and they are prone to machine learning based attacks \cite{delvaux2017}.
\vspace{-7pt}
\section{Problem Definition and Solution}
RF-fingerprinting being an up-and-coming solution for the Internet of Things has certain well-established drawbacks :\vspace{-14pt}
\subsection{ Reliability vs. Vulnerability} 
RFFs are either too feeble to survive the wireless channel, or the variations are too simple or obvious and can be easily attacked.   There exists a trade-off between reliability and security \cite{Robyns2017}, mainly due to the wireless channel.  A high spoofing resistance means that the system has a low noise resistance. Conversely, if an RFF is designed to detect even the most feeble variations, its reliability takes a hit, as the channel variations make the legitimate devices unrecognizable.\vspace{-7pt}
\subsection{Weak Distiguishability and Reduced User Capacity}
Typical experimental validations of  RF-fingerprinting use either high-end measurement device (like software-defined radio \cite{Fran19}) or network analyzer) or use a line-of-sight (LOS) channel with high SNR \cite{Brik2008}. When low-cost off-shelf devices are used (transmitter and receiver) in a noisy multipath channel, the distinguishability of the authenticating devices is greatly reduced \cite{Wang2017}. This leads to the reduction of the user capacity of the RFFs.\vspace{-7pt}
\subsection{The Complication in Data Processing}
Limited distinguishability and adverse effect of the channel leads to the usage of complicated data processing \cite{Zheng18} or large time series data \cite{Wang2017}, including complex feature extraction mechanism for mining the RFFs.  Noise and multipath components in the real world channel also adds to the complexity of RFF mining.  Recently, an attempt to minimize the channel effects was proposed with the use of tunable FIR filter at the transmitter \cite{Fran19}.  This FIR tuning required computation of the filter taps at the server, which are then downlinked back to the transmitter to improve the RFF accuracy. Although this improved the reliability, it increased the hardware and power cost of using a continuously tunable FIR filter in the hardware and a 3 step authentication procedure having security flaws.\vspace{-7pt}
\subsection{Design Consideration: Need for Configurability}
Let us consider the mathematical model of an RF-Fingerprinting system. For an IoT network with $N$ nodes, let the $i^{th}$ node be transmitting a signal to the server. The received signal $r(t)$ received by the server can be formulated as: 
\begin{equation}
r(t) = F_i\big[x(t)\big] \ast h(t)+\eta(t),
\label{eq:y}
\end{equation}
where $F_i[\cdot]$ is the transceiver effect imposed on user $i$, $x(t)$ is the transmitted signal, $h(t)$ is the wireless channel, and $\eta(t)$ is the noise. These features of each transceiver, i.e., $F_i[\cdot]$, are unique RFFs and can be observed from the electromagnetic waves that are emitted by it.  However, with developing manufacturing processes, these intrinsic variations are diminishing. The variation caused due to the function  $F_i[\cdot]$ diminishes and becomes indistinguishable from that of the other node, given practical environmental noise. This greatly reduces the user capacity of the RF-Fingerprinting system. Furthermore, if the threshold conditions are adjusted to accommodate such a minute variation, it would make the system unreliable.
\subsubsection*{Design Requirements: }
\begin{enumerate}[a)]
\item \textbf{Configurability:} A discernible solution to the above problems, is to design the intrinsic variation $F_i[\cdot]$, which would ensure that the patterns are different for different users. It can also be perceived as intentionally creating some perturbations in the transmitted signal unique to the IoT device.  If $F_i[\cdot]$ is configured, a balance between reliability and security can be designed based on the wireless channel. It can also eliminate the limited distinguishability and the need for complicated data mining since the features are software controlled.
\item \textbf{Channel Robustness:}  The design has to allow the RFF to be prominent through the wireless channel.
\item \textbf{Low power consumption:} IoT devices with batteries are power-constrained and cannot expend much energy for authentication alone.
\item \textbf{Non-intrusiveness and integrable:} This solution should neither adversely affect the communication link not complicate the protocol in use.
\item \textbf{Cost-effectiveness:}: Several IoT applications cannot afford expensive hardware.
\end{enumerate}\vspace{-8pt}
\subsection{Proposed Solution}
In this paper, we propose to use `metasurface' (A brief introduction about metasurface is given in section \ref{design}) to inject a controlled, well-intended RF Fingerprint, which is more prominent and can carry more complex features in the physical layer economically. Using the metasurface (Intelligent reflective surface) we create constructive signal additions at different frequencies within the same band. Our system is very energy efficient as it can work with no power or with few hundred micro-joules depending on the application.  Furthermore, It can be built cheaply for $\$ 0.2$ or less. Our  solution also makes the signature injection part of the transmitter non-intrusive, i.e.,  Our metasurface can be attached to the transmitter (IoT device) and is designed to induce electromagnetic changes in the waves that are emitted by the transmitter (an example application would be cellphone case or additional inexpensive cover for the IoT device). 
Since this is attached at the transmitter end, the function $F_i[\cdot]$ would be a convolution of the designed property induced by our metasurface $s(t)$. 
\begin{figure}[b]\vspace{-16pt}
    \centering 
    \includegraphics[scale=0.33]{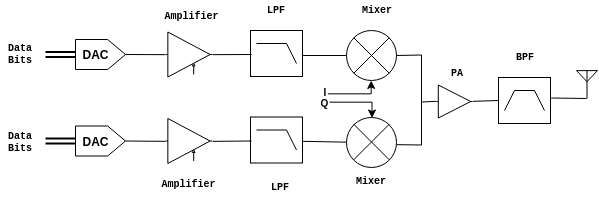}
    \caption{Typical Direct Conversion transmitter}
    \label{fig:Txblock}
\end{figure}\vspace{-7pt}
\subsection{Analysing IoT hardware for RFF injection} 
Since our solution is a low cost accessary to the IoT devices aiding in its authentication, it is necessary to analyze the hardware components of the IoT transmitter with the objective to find the answer to two important questions. The primary one is 

\textit{a) Can a security signature be injected into the wireless physical layer by changing the existing hardware? } 

To answer this question let us consider the injection of such a signature by making changes specifically in the passband of the transmitted signal.  In a typical IoT transmitter such as the direct conversion transmitter \cite{choi2014} shown in  figure \ref{fig:Txblock}, the digital to analog converter (DAC's)takes in data bits as in-phase and quadrature components. The output of the DAC is then filtered and sent to a mixer, where the baseband is up-converted to RF frequency. The Mixers are driven I and Q local oscillator inputs. The output is then passed through a power amplifier (PA) and is sent to the antenna drive circuit via a bandpass filter (BPF reduces out of band emissions). All the existing RFFS are created due to the imperfections in the above-mentioned components \cite{Danev2012}.  

To create controlled perturbations of any of these respective components is not easy.  Power Amplifiers of IoT transmitters are carefully designed with the tradeoff between nonlinearity and power-efficiency \cite{Edalat03}. The power amplifier distortions, even though are hardware signatures, can result in a performance loss; hence, they are carefully designed in a certain operating point \cite{Edalat03}. Thus designing a variation in PAs is too complicated, given the constraints of it affecting the performance.  Similarly, reconfigurable mixers \cite{Circa05} do exist, but these reconfigurations do not create distortions. Mixer distortions would create synchronization and timing issues, thus affect communication performance.  The DAC's are also very carefully designed with complex constraints to reduce the power for IoT transmitters \cite{Khorami2018}. Thus introducing configurable distortions to these transmitter entities may prove very costly, which leaves us with the RF filters.  Injection of an RFF can be made possible via a digital filtering technique, which would create that feeble variation or make deliberate changes to the hardware that would create uniquely identifiable signatures from device to device. Thus our questions on the feasibility of injecting RFF is narrowed down to the possibility of creating perturbations in the wireless physical layer with the help of carrier shaping digital RF- filter.  This brings us to another important question that needs to be answered,

\textit{b) Can this signature injecting, RF-filter be cost effective and power efficient to be adopted for IoT authentication?}

Let us consider the hardware realization of a passband filter that can generate a fingerprint. We group them as two main methods of realization, which is as described below:

\subsubsection{FIR filter with window} Firstly, we can design a digital finite impulse response (FIR) filter, apply a  suitable window function, then approximate the functions to make them implementable \cite{syed16}. They can be implemented either by using a separate DSP chip \cite{Mehen98} or by designing a dedicated system on a chip (SoC) \cite{Erdo00} comprising of latches, multipliers, and adders.

\begin{enumerate}[a)]
\item \textbf{Complexity analysis:} Consider FIR filter with a Kaiser window that can alter the emitted carrier wave, which can supposedly act as an RFF.  The choice of this specific method to analyze was considered primarily because the FIR  method can be represented as integer math \cite{syed16} and thus making them realizable with hardware compared to IIR filters. Secondly, a typical kaiser window \cite{Johnson09} is applied here, which is appropriate for generating sharp peaks. This algorithm can be considered as 3 steps. Firstly, the computation of $FFT$, which has a computational complexity of $O(M)$ \cite{Johnson09}, where $M$ is the number of samples. Secondly,  the computations of FIR filter output, which has a fixed number of multiplications and additions involved. Finally, the computation of a Kaiser window, which has the complexity of $O(N*log(N)$, where $N$ is the number of frequency samples points. The implementation of the first two steps can be performed either by DSP \cite{syed16} or an SoC \cite{Erdo00} as previously mentioned and the computation of custom window needs a numerical function generator (NFG) \cite{Tim07} unit.
\item \textbf{DSP implementation:} Although there are very cheap DSP's available for as small a price as $\$2$ \cite{DSP14}, they have an idle power consumption of $0.15 mW$ at $1.05 V$ supply \cite{DSP14}. Furthermore, the delays involved in using a DSP unit is larger because it needs to perform the total filter operation in iterations. This is because the DSP relies on a single efficient multiplication and adder unit (MAC) to perform the operation per loop \cite{Serrano2008}. The resulting power consumption and time delay are not affordable for an IoT device.
\item \textbf{SoC implementation:} The same filter implemented using SoC, however, would be faster as they can perform the operation in a single flow, but the complexity of the 3 stages are tough to be designed in a single chip without significant approximations \cite{syed16}, as they rely on implementing the operation with only latches, multipliers, and adders. Also, the power consumption of such a filter to of size $N=16$ is estimated to be $28.7 mW$ for low throughput applications \cite{Moh16}, thus making them unaffordable for IoT devices.
\end{enumerate}
\subsubsection{RF resonator Filters:} Another way of making the variation could be through RF filters \cite{Alyasir2019}. Out of the RF filter technologies that exist, such as discrete inductor-capacitor (LC) filters, multilayer and monoblock ceramic filters, acoustic filters, and cavity filters,  the current state of the art wireless communication chips employ BAW filters \cite{Ruppel17} to achieve operation.  Even though devices with higher Q-factor and lower insertion loss have been designed, complex signatures with configurable BAW filters \cite{Hashi11} of RFF capability are a few decades away from becoming a reality. It is difficult to design these filters with precise cutoff, group delay, selectivity, ripple, and isolation between input and output signals \cite{RF15}. 

From the above discussion, it is clear that making the transmitter chip with reconfigurable hardware components for signature injection would prove to be costly, inefficient and may affect communication performance.\vspace{-7pt}
\subsection{Configuration for Different Security Level}
We present two types of metasurfaces that can be employed with the design goals we have set.
\begin{enumerate}[a)]
\item \textbf{Passive metasurface} - This metasurface has a well designed, feature rich signature injection capability but is static. Thus making it suitable for very low-cost applications.
\item \textbf{Active (programmable) metasurface} - This metasurface can change its property based on a control input while maintaining the richness in the feature space. This feature is very suitable for IoT applications, which are not strictly energy constrained but needs a more secure solution as the injected signature can change with respect to control input.
\end{enumerate}
With the capabilities as mentioned above, we designed and implemented the metasurface RF fingerprint injection (MeRFFI) System, which is a programmable metasurface, MeRFFI prototype has the capability to work in both the modes due to its programmability.

We present three different ways in which MeRFFI can be used for IoT authentication.
\subsubsection{Passive Mode}\vspace{-2pt}
\textbf{Protocol 1 - low power application : } Here the MeRFFI Device used does not have channel selection or signature injection control mechanism. It is a passive meta-surface like a passive RFID card, which consumes no additional power. A different pattern is printed on MeRFFI, and these signatures are injected into the wireless physical layer. During authentication, the IoT node (I) sends an authentication request signal (Req) along with its device id (DevID) to the access point (A). This signal is injected with the physical signature (IRFF1) from MeRFFI.
\begin{equation*}
    I \rightarrow A : Pilot\{Req,DevID\}|IRFF1
\end{equation*}
The access point (A) extracts the injected channel state information
($inj\_C\_Est$). It forwards this request ($A_{Req}$) to the authentication server ($S$).
\begin{equation*}
    A \rightarrow S : A\_Req\{inj\_C\_Est,DevID\}
\end{equation*}
The authentication server extracts the features ($fp$) and checks for a match. If the feature and DevID match to that of the registered device, the server sends back an acknowledgment message with a hash code of the feature ($H(fp)$) along with the target DevID. When node $I$ receives this authentication acknowledgment, it can also verify that it is associating with the correct server from the hash $H(fp)$.

\subsubsection {Active Mode}\vspace{5pt}
\textbf{Protocol 2 - highly secure application and high power consumption : } This protocol uses the active version of MeRFFI that uses control voltage based signature injection. The IoT node ($I$) sends a request to the authentication server ($A$) using the same message flow as in the previous protocol. The authentication sends back the acknowledgment along with the hash of the feature extracted from the pilot signal ($H(fp)$). On receiving the $A_{Req}$ the IoT node $I$ does a server hash correlation and verifies the hash. After the hash is verified the node pushes $n-1$ number of designated signatures, which are allotted to the node. On receiving all the signatures, the authentication server correlates all the received signatures ($IRFF1, IRFF2,....IRFFn$) with the device id (DevID). MeRFFI's high user capacity supports this highly secure protocol. \vspace{5pt}   

\textbf{Protocol 3 - highly secure application and medium power consumption :} 
In this protocol, multiple transmissions are made similar to protocol 2, but the property of the metasurface is not changed. Here the injected signature on the IoT node (I) is fixed, and transmissions are made from different channels to the authentication server (A). This protocol is suggested based on the hypothesis that the same signature appears as distinct from different channels. This is later validated with our experiments. The difference in the property of MeRFFI in different channels is leveraged here to create a highly secure protocol. The power consumption here would be less than protocol 2 since there is no need to switch the metasurface.
\begin{figure*}[h t p]
    \centering
    \includegraphics[height= 1.2 in, width=6.1 in ]{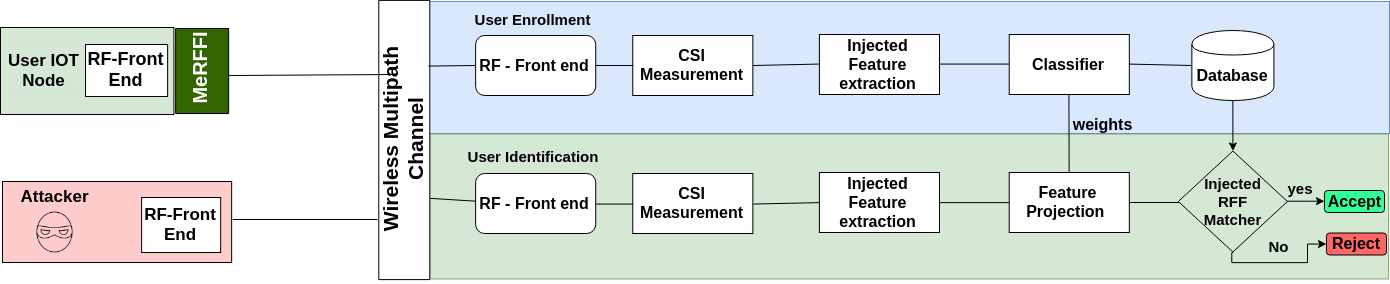}
    \caption{System model: MeRFFI based user enrollment and authentication system. Unauthorized user or attacker will be rejected by the system.}
    \label{fig:Sysmodel1}
    \vspace{-10pt}
\end{figure*}
\begin{figure}[t]
    \centering
    \includegraphics[scale =0.18]{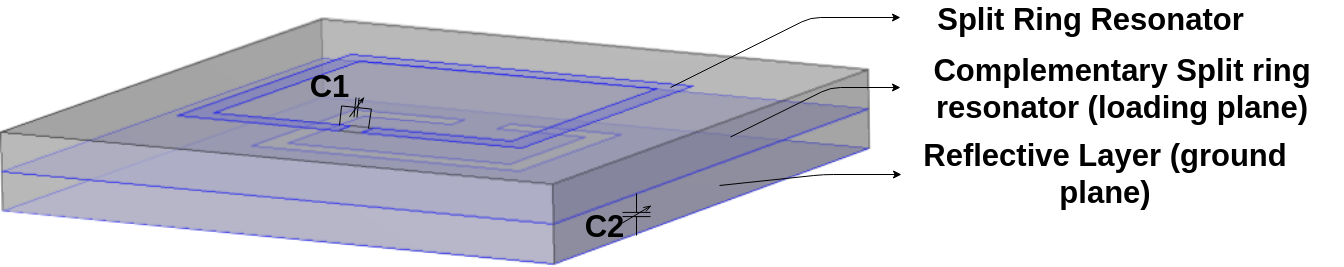}
    \caption{Geometry of a single miniaturized meta-material element used to make the meta-surface}
    \label{Metadig}
    \vspace{-8pt}
\end{figure}
\begin{figure}[ht]
\includegraphics[scale =0.42]{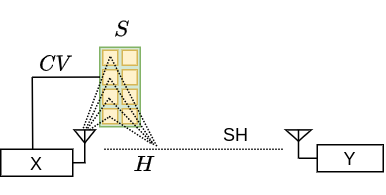}
	\caption{Working model of MeRFFI}
	\label{fig:MeRFFIsysrep}
	\vspace{-10pt}
\end{figure}\vspace{-8pt}
\section{Meta-surface RF-Fingerprint Injection (MeRFFI) System Design}\label{design}
From section
The radio frequency fingerprint injection system is described along with the basic user enrollment and authentication scheme. Consider a scenario as shown in figure \ref{fig:simplepic} where there are $N$ nodes trying to connect with the authentication server. All the registered nodes in the network are attached with the MeRFFI. As it can been seen in the figure \ref{fig:Sysmodel1}, during the enrollment phase, the legitimate node sends a known pilot signal to the server over the wireless channel. The attached MeRFFI device injects an RF security signature in this transmitted electromagnetic wave. The server on receiving the node's signal measures the channel state information (CSI) and then extracts the injected features from the measured CSI. A classifier trains and classifies the signatures of all the legit user devices and stores them in a database. During the authentication phase, when the user nodes request access, the server measures the CSI and extracts the features of the signature from this request signal. The projected features are then matched with the enrolled user devices and is granted authentication on a match. An attacker node, on the other hand, would not have the injected signatures; hence their authentication will be denied.

MeRFFI knows the communication channel number (specific frequency) allocated for communication. It can tune to the present channel of operation. Then it supplies a software programmed voltage vector input, where all unit elements are controlled by each in the voltage values. This results in MeRFFI altering the wireless physical layer by injecting a physical signature.\vspace{-7pt}
\subsection{Metasurface Preliminary}
Metasurfaces \cite{liu2018} are ultra-thin and planar electromagnetic devices, which are made up of several unit elements. These unit elements are sub-wavelength in size compared to traditional microwave patches and printed antenna elements. They have heavily gained popularity because of the tunability of their individual unit elements \cite{stanislav2016}.

The metasurface for RF-fingerprint injection is a prototype that has \textit{software programmed tunability}. The material design specification and its tuning control are described below.

\begin{figure*}
\includegraphics[width=2\columnwidth]{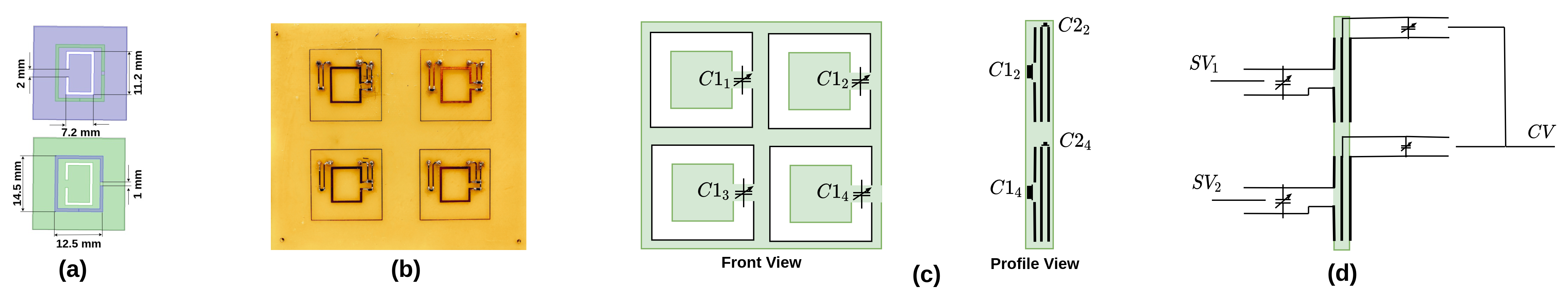}
	\caption{MeRFFI Prototype:(a) Dimension of the top layer (SRR) and the loading plane (CSRR) (b) Four unit prototype, 2 such boards were used in our experiments (c) Front View and side view of meRFFI's 4 unit board, showing the varactors (d) control mechanism}
	\label{fig:MeRFFI_all}
	\vspace{-8pt}
\end{figure*}\vspace{-7pt}
\subsection{MeRFFI Prototype Design}
MeRFFI is a software-controlled metasurface used for injecting the security signature. It is an Intelligent Reflective Surface (IRS), a.k.a Reconfigurable Reflect Array, which has many unit cell structures \cite{Ozdo20}. The typical IRS's unit cells can change their resonant frequencies, which can be perceived as a change in its reflectance and phase.  Our custom designed metasurface, MeRFFI, creates constructive reflections on the signals transmitted.  These small perturbations appears as frequency peaks when observed from the channel state information at the receiver.  To design this specific reflective surface, the unit element has two main requirements. Firstly, to bring the entire metasurface's resonant frequency to the specific frequency channel where the transmission is taking place. Secondly, some finer modifications of resonances within this channel cause different electromagnetic variations in the wireless physical layer.

\textbf{MeRFFI's unit cell structure:} MeRFFI prototype's unit cell has three layers: 1) a frequency selective resonant layer; 2) loading plane, which helps in miniaturizing the dimensions of the metamaterial, and; 3) a reflective ground plane. The top layer is a split ring resonator (SRR), which at the designated frequency, reflects an incoming wave. There is a capacitor $C1$ connected between the slit after split-ring resonator in top layer to change the resonant frequency this SRR. The second layer is a complementary split ring resonator (CSRR) used to miniaturize the unit element\cite{Oloumi12, Tang2014}. Finally, between the there is a variable capacitor $C2$ connected for changing larger values of resonance.  \\
\textit{Specifications}: The meta-surface unit element design is shown in figure \ref{Metadig}. We have used a split ring resonator as the top plane. A complimentary split ring resonator is used as the loading plane \cite{Meng20, Bayatpur2009}. The dimensions of these layers are given in figure \ref{fig:MeRFFI_all}(a). We have used varactor $smv2023-079LF$ \cite{Sheet2015} as C1 and C2. This lab prototype is shrunk to a half-wavelength dimension of its original size, but for usage in much smaller devices, this can be miniaturized further \cite{Tang2014}. The entire unit element is built using an FR4 substrate, which is a common PCB material.

\begin{figure}[h]
	{\includegraphics[width=2.6in ,height=1.6in]{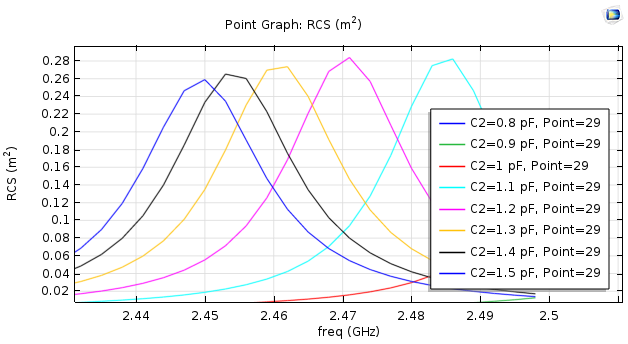}}
	\caption[Caption for LOF]{Variation of Unit cell resonance with change in capacitance}
	\label{RCS}\vspace{-17pt}
\end{figure}

\textbf{MeRFFI's control mechanism :} Figures \ref{fig:MeRFFI_all}(c) and \ref{fig:MeRFFI_all}(d) shows the voltage inputs given to  all the varactors in MeRFFI prototype. By changing the input voltage to the capacitor $C2$, the capacitance between the loading plane and reflecting plane changes. Since this control voltage ($CV$) is connected to all the unit cells, the total resonance of the meta-surface shifts to the desired frequency channel. Hence this voltage $CV$ is the channel select voltage. The change in resonant frequency with variation in capacitance is shown in figure \ref{RCS} (The graph shown is a Comsol simulation of MeRFFI, where the capacitance is changed by control voltages, the value displayed here is frequency vs. radar cross section [RCS($m^2$)])
The other varactors $C1$ on the top layer, however, have independent voltage connections. If MeRFFI has $M$ number of elements, then an $M$ line parallel but independent control voltage line is needed for controlling the varactors ($SV_1,SV_2,.....SV_M)$. This $M$ line control voltage is what can be used to inject RF-fingerprints into the wireless physical layer. The MeRFFI prototype has 4 unit elements as shown in Figure \ref{fig:MeRFFI_all} and two such prototypes were used for signature injection. Hence MeRFFI prototype had an 8 line control input vector. The metasurface prototype has eight such elements in which an 8-line control voltage feed can be given for creating many such variations.\vspace{-6pt}

\subsection{Mathematical Modeling}\vspace{-1pt}
For modeling the injection of a physical layer signature, consider the $i^{th}$ transmitting IoT node which has MeRFFI is placed in its proximity, such that any emission from the $TX$ node reflects from the MeRFFI as shown in the figure \ref{fig:MeRFFIsysrep}.  Let $x(t)$  be the signal transmitted by the $i^{th}$ IoT node ($X$ in the frequency domain) and  $s$ is the function of the  meta-surface. Any transmission $x$ going out from this $TX$ node would be the convoluted with $s$ as it is emitted \cite{Ozdo20}. if $h$ represents the multipath channel between node $i$ and the authentication server, the expression for received signal $y$ can be written as,
\begin{equation}
y(t)=x(t)*s(t)*h(t)+\eta,
\label{eq:rxsignal}
\end{equation}
The function $s$ can be considered as the sum of the reflection from individual unit elements whose magnitude and phase is controlled by the control voltage vector $CV$. The fourier transform of $s$ which has $N$ unit elements can be written as,
\begin{equation}
S=\sum_{n=0}^{N-1}\rho_n e^{j\phi_n-2\pi f\tau_n}
\end{equation}
where $\rho_n$ and $\phi_n$ are the magnitude and phase of the response from  the $n^{th}$ unit element, and $\tau_n$ is the time delay of the $n^{th}$ impulse in the time domain. The channel response received at the receiver as seen in equation \ref{eq:rxsignal} is a convolution of the MeRFFI's transfer function ($s$) and the channel multipath ($h$) similar to the sum of the product of amplitude and phase responses caused by the individual path in the multipath, the fingerprint is also the sum of the effect caused by the individual unit elements in the reflect array \cite{Ozdo20}. 
In the frequency domain this injected channel response $SH$, is given as:
\begin{equation}\label{eq:4}
SH_j=\sum_{m=0}^{M-1} \sum_{n=0}^{N-1} \alpha_m \rho_n e^{j(\theta_m+\phi_n-2\pi f[\tau_m+\tau_n]  )} ,
\end{equation}
where $SH_j$ represents the $j^{th}$ frequency sample of the signature injected channel and it is a complex number that includes the amplitude $|S||H_j|$ and the phase $\angle{(\phi_{h_j+}\theta_s)}$ is shown as follows.
\begin{equation}
SH_i = |S||H_j| \angle{(\phi_{h_j+}\theta_s)},
\end{equation}
Thus at the receiver we get the injected channel vector
$SH$ for $s$, CSI samples (here, the number of CSI samples $j=s$) , i.e ,
\begin{equation}\label{eq:SH}
SH_i = [SH_1, SH_2, SH_3, . SH_j, . . . , SH_s]^T, 
\end{equation}
In general, 
\begin{equation}
Y = SHX + \eta,
\end{equation}

From the received signal vector $Y$ we extract $SH$. The fingerprint induced CSI of the IoT node to the authentication server has to be extracted.
\begin{equation}
SH=\frac{Y-\eta}{X},
\end{equation}
After obtaining $SH$ the injected fingerprint has to be separated from this vector. Thus we can define a feature extraction space $\chi$, where the feature of a single fingerprint is translated to an N-dimensional feature space.          
\begin{equation}
\chi:SH\rightarrow F\approx \hat{S},
\end{equation} 
where $F = \{f_1,f_2,f_3...f_N\}$. After this we find a function $\Omega$, that can project the feature space of $N$ users, $F$, to class space, $C = \{c_1,c_2,c_3....c_N\}$, i.e.,
\begin{equation}
\Omega:F\rightarrow C,
\end{equation} 
where the function $\Omega$ belongs to a hypothesis space $\omega$.
After the projection, the feature of the specific user is classified to the corresponding class.\vspace{-6pt}
\subsection{Signature Extraction and Classification}\vspace{-3pt}
 As mentioned in equation \ref{eq:SH}, we receive a vector of complex values corresponding to the number of CSI samples in the communication channel.
A convolutional neural network (CNN) was used to extract the signature from the computed CSI vector $SH$. Although CNN is more computationally complex, a server in IoT can afford the computational expenditure during user enrollment, the matching (authentication) phase would be just as simple as any other classfier algorithm. In MeRFFI's implementation, the first three layers of the CNN are used to extract features. In each layer of the CNN, one-dimensional kernels were used as the filters. This was followed by a batch norm layer to normalize the mean and variance of the data at each layer. For the third layer, a rectified linear unit (ReLU) was added to introduce nonlinearity and a max-pooling layer to reduce the size of the representation.If $\chi$ is the set of parameters for feature extraction in the CNN, then we get the feature $F$ as:
\begin{equation}
F = CNN(SH;\chi).
\end{equation}
Depending  on the outputs of feature extractor (i.e., F), a fully connected layer followed by another ReLU is used to learn the representation $K$of $F$ as follows:
\begin{equation}
Ki = W_f f_i + b_f,
\end{equation}
where $W_f$ and $b_f$ are the parameters to be learned and the softplus function is an activation function to introduce nonlinearity,
in order to identify the injected signature,  the feature representation $K_i$ into a new latent space $H_i \in {\rm I\!R_m}$, where $m$ is the total number of signatures to be classified.
Finally, a \textit{Softmax} layer is used to obtain the probability
vector of signatures:
\begin{equation}
C_i = \textit{Softmax}(H_i),
\end{equation}\vspace{-15pt}
\begin{equation}
H_i = W_k K_i + b_k,
\end{equation}
where $W_k$ and $b_k$ are parameters. From this, we obtain classes corresponding to the signatures $C_i$. A loss function $L(c,\hat{c})$ can be defined where $\hat{c}$ is the estimated class. The expected loss of $\Omega$ (the entire classification function) can be estimated while training, using an empirical risk function $R$, where 
\begin{equation}
R= \frac{1}{N}\sum_{i} (c_i,\Omega(v_i)).
\end{equation} 
This measure helps us in avoiding overfitting and underfitting scenarios while training. After the classifier is trained, the classified features of the enrolled users are then stored in a database. When a new incoming signal is received, its feature is extracted and its score is computed with the features saved in the database, which is then used for authentication.\vspace{-15pt}
\subsection{Reconciling MeRFFI to Channel Dynamics}\label{cdyna}
To design the MeRFFI metasurface to inject a robust RFF into the wireless channel, we need to take in to account the dynamics of the channel. Let's consider a scenario where node $i$ is at a location $x\in \mathbb{R}^2$ and the receiving base station at  $x_b\in \mathbb{R}^2$. The received signal strength $\gamma(x)$ can be represented as a 2D non-stationary field made up of three types of channel dynamics \cite{JRobotics11}; path loss ($\gamma_p(x)$), shadow fading ($\gamma_s(x)$) and the multipath fading component ($\gamma_m(x)$).
\begin{equation}
\gamma(x)=\gamma_p(x) \gamma_s(x) \gamma_m(x)
\end{equation}
Let the injected the injected signature be represented by the random variable $\gamma_{sig}(\vec{SV},CV)$, which is dependent only on the channel select voltage $CV$ and signature inject voltage vector $\vec{SV}$ , the received signal strength becomes,
\begin{equation}
\gamma(x)=\gamma_p(x) \gamma_s(x) \gamma_m(x) \gamma_{sig}(\vec{SV},CV)
\end{equation}
In Decibels this equation is,
\begin{equation}
\begin{aligned}
\gamma_{db}(x)=10\log(\gamma_p(x))+10log(\gamma_s(x))+10log(\gamma_m(x)\\+10\log(\gamma_{sig}(CV,\vec{SV}))
\end{aligned}
\label{eq:pgain}
\end{equation}
 From various indoor channel modeling based on real-world measurements \cite{JRobotics11} , $\gamma_p(x)$ is a distance dependent path loss, and $\log(\gamma_s(x))$ follows a zero mean gaussian distribution and the channel multipath $\gamma_m(x)$ follows a rician distribution as given by,
\begin{equation}
f_w(x)=(1 + K_r)e^{-k_r-(1+k_r)x}I_0(2\sqrt{xk_r(k_r+1)}
\end{equation}
Where $K_r$ is the Rician K factor, which is the ratio between the power in the direct-path ($p_d$) and the diffused power, and $I_0(.)$ is the $0^{th}$ order Bessel function.
By applying the expectation operator on equation \ref{eq:pgain} and rearrange it to inequalities we get,
\begin{equation}
\begin{aligned}
\gamma_{db}(x)-10\log(\gamma_{sig}(CV,\vec{SV})>10\log(\gamma_p(x))+\\10log(E(f_w(x)))
\end{aligned}
\label{eq:pgain2}
\end{equation}
Here the path loss has zero mean and we can deduce 
that the gain of the RF-fingerprint injected should be greater that the multipath power.

Consider a  dynamic environment of an area of radius $||x-x_b||$ with $N_{obj}$ moving objects each at distances $r_i$ of the receiver and having an average width of $\delta w$. These objects absorb some part of the energy, by an absorb factor $\zeta$ and the total path power is $P_m$. If the objects are uniformly distributed with in the area has the following $k_r$ \cite{Ohta08}, 
\begin{equation}
k_r=\frac{p_d}{p_m}\frac{\pi(r_{max}+r_{min})}{N_{obj}|\zeta|^2\delta w}
\end{equation}
From this equation, we can infer that the rician factor and hence mean dynamic value of the multipath power is influenced by the number of moving objects per unit area \cite{Ohta08} ($K_r= 0$ means a highly dynamic channel). 
\subsection{Channel Robust Metasurface Design}For the injected signature to be prominent in this dynamic environment, MeRFFI metasurface's SNR with in a short delay spread has to be above the threshold level of the maximum variations caused by channel dynamics. Practically this is the metasurface's expected capability where it provides constructive interference on the channel, which aids the communication by enhancing certain narrow frequency components. This is contrary to the frequency selective fading introduced by the multipath and requires the metasurface to be tuned to the operating frequency. The maximum gain of the metasurface is obtained when the perimeter of the unit cell is of the order of the wavelength used for communication ($\lambda$). The gain reduces when the unit cell's size is shrunk. The idea is to choose the metasurface's size to make the constructively added perturbations to withstand channel variations and at the same time conform to the devices form factor. Our prototype is thus built with the perimeter of each unit cell of the order of $\lambda/4$ (65 mm), which is sufficient for a time varying, dynamic, real-world indoor environment.

\textit{Remark:} Designing the metasurface channel robust also has the advantage that the injected Rf-fingerprint dataset size becomes independent of the channel, robust to time variance and depends only on the injected code. \vspace{-6pt}
\section{Experiment and Results} \vspace{-3pt}
As indicated in the previous sections, MeRFFI is a low cost and small attachment to the IoT transmitter's hardware that can inject a radio frequency fingerprint (RFF) into the wireless physical layer. Technically, MeRFFI's hardware can be used either in passive mode, where it injects a fixed physical signature, or in active mode, where it can inject varying physical signature according to an input control voltage vector and also has the capability to select the channel (frequency) of operation. To test MeRFFI's performance both in active and passive mode, the prototype as shown in figure \ref{fig:MeRFFI_all}(b) was built in the lab. Extensive experiments have been performed using this MeRFFI prototype for testing its potency as a wireless physical layer injection system for the IoT applications. All the experiments were performed in a dynamic environment wherein between 2 and 6 people are always present in the testbed's vicinity. The testing duration also spanned between 4 and 12 hours per day over 60 days to give the collected dataset a time varying factor.

In this section, we present MeRFFI's control mechanism set up, performance measures used followed by the experiments in which the MeRFFI system was tested.  Initially, we test MeRFFI's control mechanism via an experimental setup using a Vector Network Analyzer (VNA), where we inject different control codes and visually observe the changes in the wireless channel. After making sure the proof-of-concept software programmability of MeRFFI works, we conduct our experiments on a WiFi commercial off the shelf (COTS) testbed to test MeRFFI. Our COTS testbed utilizes laptops Lenovo ThinkPad E570 and Dell  Latitude  E6530 with Intel  5300ac  NIC card  to replicate the different static IoT deployment scenarios and check whether MeRFFI satisfies the specified design goals:

(a) Inject 208 closely-spaced control codes and observe the performance to estimate the user capacity of our prototype.
(b) Test the performance with change in distance. Here, changes in distance upto 53 m in an indoor multipath channel were made and the resulting performance was analyzed.
(c) Test the impact of orientations of the user devices.
(d) Test the effect of blockage by the through-the-wall experiments.
(e) Test the performance of the system, when same set of signatures are injected in different WiFi channels.\vspace{-7pt}
\subsection{MeRFFI Metasurface and Control Mechanism}
Two 4-unit cell based structures were used in all the experiments(figure \ref{fig:MeRFFI_all}). As mentioned in section 5, two types of control (figure \ref{fig:MeRFFI_all}(c) \& (d)) are required. $a)$ control voltage $CV$, which is a single parallel voltage to be supplied to all the individual elements to select the channel of operation, $b)$ control voltage vector $\vec{SV}=\{SV_1,SV_2,.....SV_M\}$, is used to inject different RF-signatures in the selected channel. We use an Arduino Mega 2560 to drive the DAC circuit to provide $CV$, and two similar sources were used to feed in the control vector $\vec{SV}$. 

Different $SV_i$ inputs correspond to an individual unit cell control input, and this was made sure with the above mentioned voltage sources and voltage divider circuits.\vspace{-10pt}
\subsection{Performance Measures}\vspace{-2pt}
For computing the performance of the classification system designed, we first compute the following: 
a) true positive (TP): the digital signatures correctly identified, b) true negative (TN): the fingerprints correctly rejected, c) false positive (FP): the false signature wrongly identified, and d) false negative (FN): a valid signature misclassified as false.  These values are used to compute the standard performance measures sensitivity ( recall, hit rate, or true positive rate (TPR), Fall-out or false positive rate (FPR), Precision or positive predictive value (PPV), Accuracy (ACC), F1 Score,  area under the ROC (receiver operating characteristic) curve.
Since this is a multi-class classification problem. We compute the micro and macro average of all the performance measures shown above. Hence we perform an $M$ number of one vs. all classifications for all the classes and compute this micro and macro averages performance measure, which would give us an overall performance of the signatures classified at the receiver. A sample of how micro and macro averages are computed. An example for computing the micro and macro average of precision ($PPV_{micro}$ and $PPV_{macro}$) is given as\\
\begin{equation}
\begin{aligned}
PPV_{micro} = \frac{TP1+TP2}{TP1+TP2+FP1+FP2}.\\
\end{aligned}
\end{equation}
\begin{equation}
\begin{aligned}
PPV_{macro} =\frac{P1+P2}{2}.\\
\end{aligned}
\end{equation}

\begin{figure}[t]
    {\includegraphics[width=1.69in ,height=1.1in]{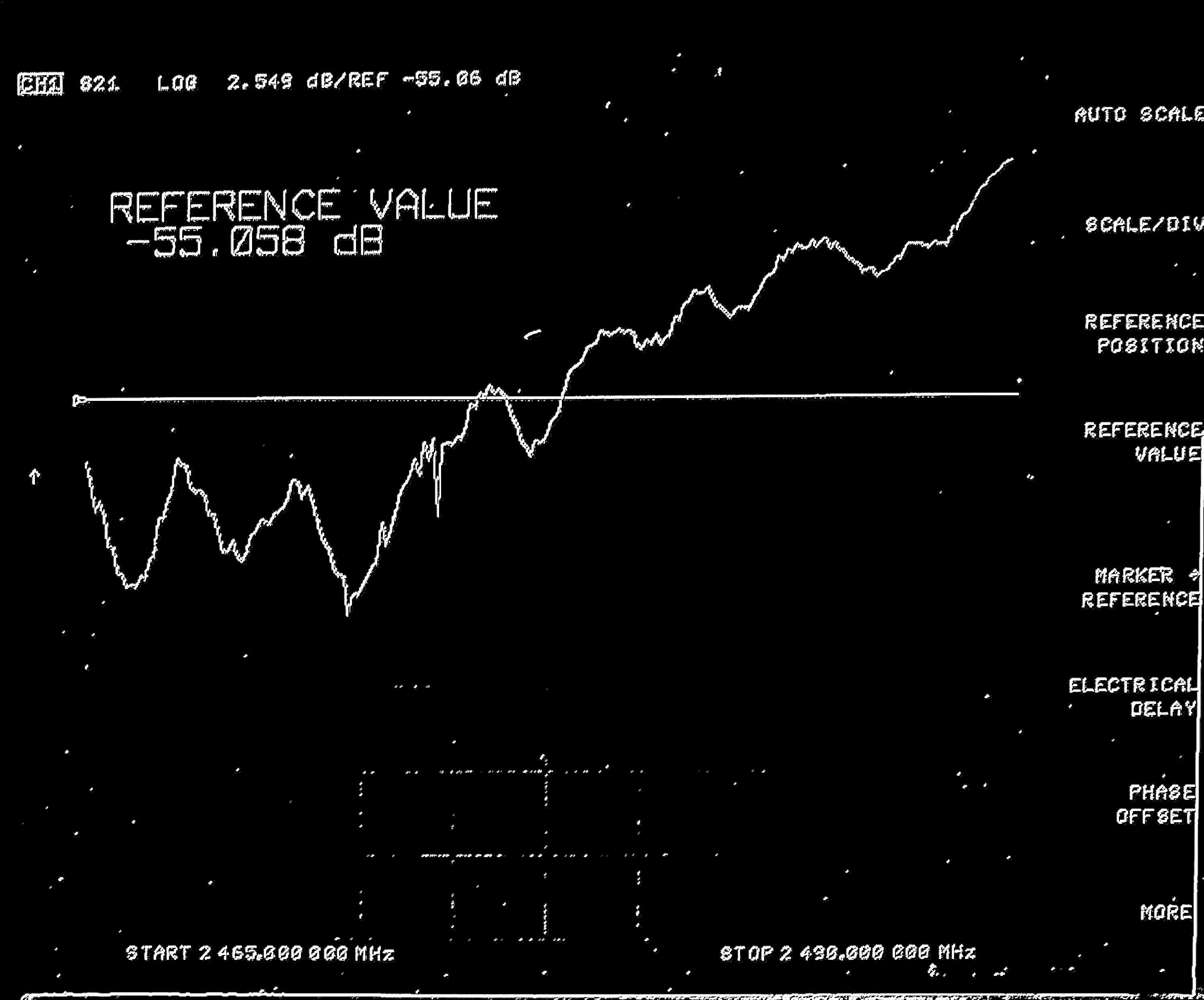}}
    {\includegraphics[width=1.69in ,height=1.1in]{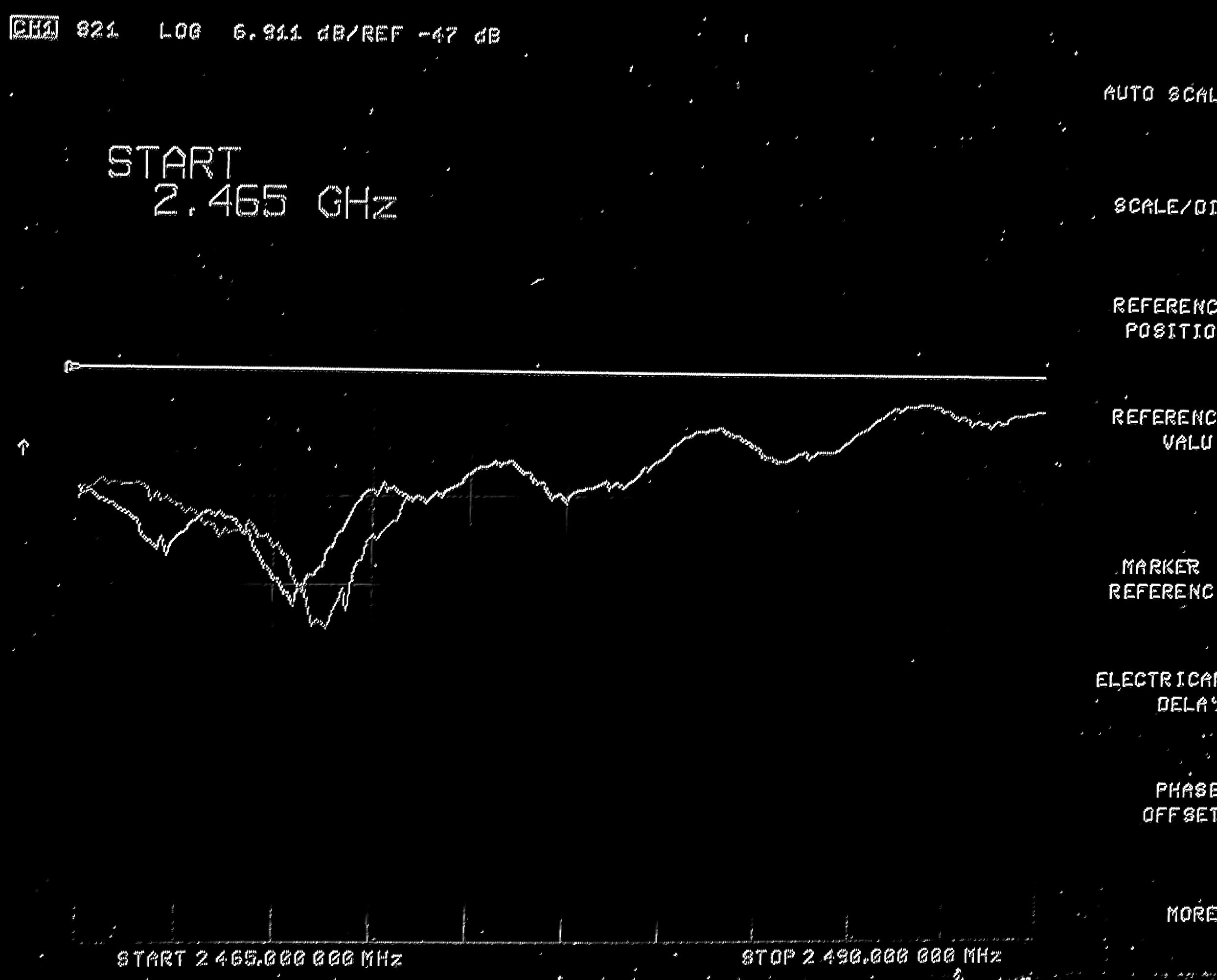}}\\
        \vspace{-8pt}
    \caption{Different patterns of forward gain $S21$ observed on the Vector Network Analyzer to different $C1$ input voltage vectors}
    \label{VNA}
\end{figure}

We use the micro and macro averages to create the average ROC and then compute the $AUC_{micro}$, $AUC_{Macro}$ average, which would give us a good sense of the overall performance of the system.
\vspace{-12pt}
\subsection{Testing MeRFFI's Control Mechanism}

\textbf{Objective:} To visually observe changes caused in the wireless channel caused by the control voltage input given to MeRFFI and thereby verifying if MeRFFI's channel select and signature inject control mechanisms are working.

\textbf{Experimental setup:} To test whether the designed meta-surface performs as per the design specification, an Agilent Vector Network Analyzer (model-8753ES) was used. MeRFFI was kept near the TX antenna, and $S21$ (forward gain) was observed from the receiver end, which is kept 2.54 m away. At first control input $CV$ was changed, and changes were observed at the VNA. The experiment was done in a conference room with the setting as seen in figure \ref{fig:TX1}.

\textbf{Results and inferences:} At 16.1 volts a peak was observed within the range 2.420 GHz and 2.425 GHz, which indicates that $CV$ was functioning as the channel selector. Then random voltage vectors $\vec{SV}$ was applied and distinct $S21$(forward gain) patterns corresponding to different $\vec{SV}$ inputs were observed as seen in the figure \ref{VNA}. This pre-test helped us validate the channel select mechanism and different patterns observed at the VNA verify that the signature injection mechanism of MeRFFI is functional.

\begin{figure}[t]
\vspace{-5pt}
	\centering
	\includegraphics[height=2.4in]{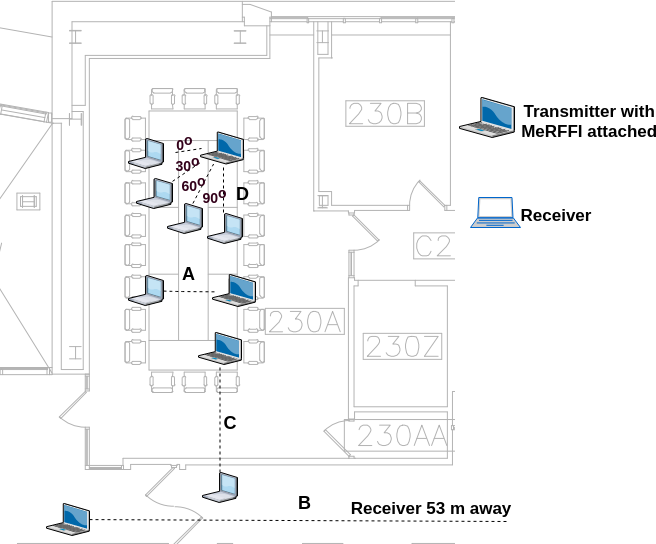}
	\caption{Room setting, where the experiments were performed.}
	\label{fig:davisplan2}
	\vspace{-14pt}
\end{figure}

\begin{table}[b]\vspace{-7pt} 
	\scriptsize
	\caption{Model Summary of the CNN used for 208 code.}
	 \vspace{0pt}
	\begin{tabular}{lll}
		Layer (Type)                & Output Shape & Param\# \\\hline
		conv1d\_1 (Conv1D)          & (30, 24)     & 120     \\
		activation\_1 (Activation)  & (30, 24)     & 0       \\
		flatten\_1 (Flatten)        & (720)        & 0       \\
		dense\_1 (Dense)            & (2048)       & 1476608 \\
		dropout\_1 (Dropout)        & (2048)       & 0       \\
		dense\_2 (Dense)            & (1024)       & 2098176 \\
		dense\_3 (Dense)            & (208)        & 213200  \\
		activation\_2 (Activation)  & (208)        & 0       \\\hline
		Total params: 3,788,104     &              &         \\
		Trainable params: 3,788,104 &              & 
        \\\hline \newline \newline
	\end{tabular}
	\label{CNN1}
	\vspace{-10pt}
\end{table}\vspace{-9pt}
\subsection{Determining Signature Capacity of MeRFFI} \label{ss:capacity}

\textbf{Objective:} In this experiment, 208 codes,  which are very close to each other (very minimum voltage change in each input control), are given as input $\textbf{SV}$. This experiment is to classify these 208 closely-spaced signatures received at the receiver and estimate the signature capacity of MeRFFI.

\begin{figure}[t]
	\centering
	\includegraphics[trim={0.1cm 0.1cm 0 0.3cm},clip,scale =0.38]{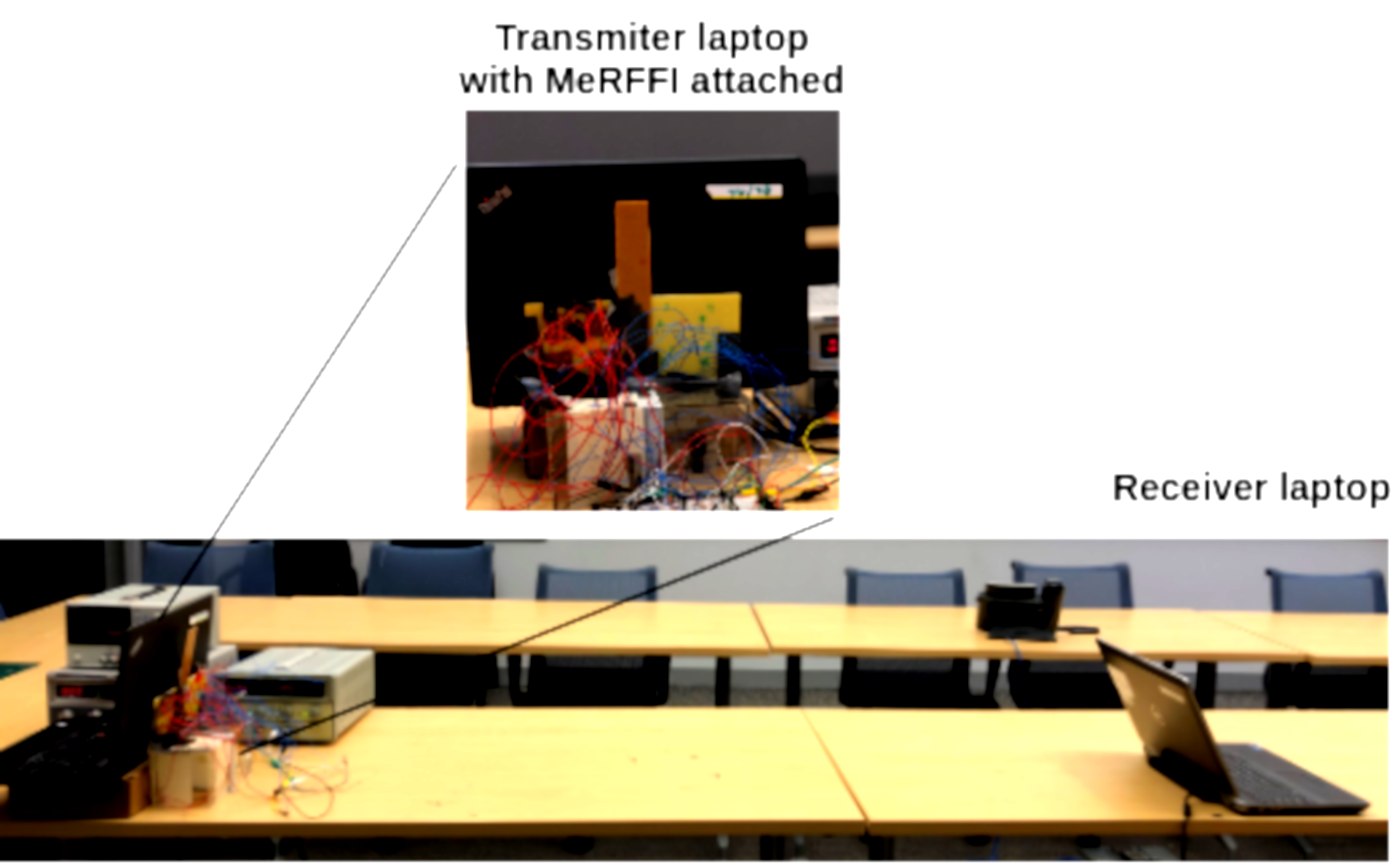}
	\caption{Experimental setup with the COTS MeRFFI.}
	\label{fig:TX1}
	\vspace{-16pt}
\end{figure}

\textbf{Experimental setup:} This experiment was performed in a conference room $10.16 \times 6.35$ m, whose plan is displayed in figure \ref{fig:davisplan2} and the position is marked with label A. This room was chosen so that the environment is not controlled, which includes other commercial wireless interferences. The COTS experimental devices used were a Lenovo ThinkPad E570 equipped with an Intel 8265ac NIC card as the transmitter, and a Dell Latitude E6530 fitted with an Intel 5300ac NIC card as the receiver. MeRFFI was placed 2 cm in front of the transmitter's inbuilt antenna as shown in figure \ref{fig:TX1}. The receiver is placed 2.54 m away from the transmitter. The values of the control voltages $SV_1$ for each unit cell are varied from 0 till 10, while the control inputs to other unit cells are kept zero. Twenty six voltage values were given to each capacitor, and the changes were done on all the eight different unit cells. Thus making it 208 different code words triggering that many injected codes, which have very minimum variation from one to the next. The number of voltage values that can be given to each unit element depends on the dynamic range of the varactor chosen. The data set collected 500 CSI vector values received for each signature. The structure of the CNN used for classifying this data is shown in table \ref{CNN1}.
\vspace{1pt}

\textbf{Result and inference:}
Table \ref{table:208t} shows the accuracy and the F1 score (micro and macro averages) for 80:20 and 90:10 test to train percentage split. The values suggest that MeRFFI RF-Fingerprinting has a substantial signature capacity and is very reliable. In the above experiment, each unit cell was triggered by 26 voltage levels one after the other.
Based on the technical specification of the varactor \cite{Sheet2015} used in creating this variation a maximum of 32 different physical signatures. 

To generalize, consider a MeRFFI prototype that has $M$ unit elements in the metasurface. Each of these unit elements can create several capacitance variations depending on the dynamic characteristics of the varactor connected to them. If $R_d$ is the dynamic range of the varactor and $V_t$ is the minimum control voltage needed to create capacitance change of the varactor, the number of signatures that can be injected by MeRFFI can be written as :
\begin{equation}
N=(R_d/V_t)^ M,
\end{equation}
 From this equation, we can interpret that a high resolution varactor with a higher dynamic range can achieve a large design space for signatures. 

In a perfect environment with low noise, the MeRFFI prototype in the lab that has 8 unit cells can create a maximum $48^8$ ($2^{45}$) signatures. Thus a more significant number of unit cells in  MeRFFI would result in a higher signature capacity.
\begin{table}[b]
\vspace{-10pt}
	\scriptsize
	\caption{Testing signature capacity, classify 208 injected signatures.\vspace{0pt}}
	\begin{tabular}{llllll}
		Test-Train\\ Split      & Accuracy & $F1_{Micro}$ & $F1_{Macro}$ & $AUC_{Macro}$ & $AUC_{Micro}$ \\\hline
		80\% test & 0.81     & 0.9848   & 0.9932   & 0.9       & 0.9       \\
		90\% test & 0.7415   & 0.9774   & 0.9781   & 0.87      & 0.87      \\\hline
	\end{tabular}
	\label{table:208t}
\end{table}
\begin{table}[b]
\vspace{-5pt}
\caption{Performance with varying distance.\vspace{0pt}}
	\scriptsize
	\begin{tabular}{llllll}
		Distance & Accuracy & $F1_{micro}$ & $F1_{macro}$ & $AUC_{micro}$ & $AUC_{macro}$ \\ \hline
		7 m      & 0.8914   & 0.9819    & 0.9407    & 0.9407     & 0.9412     \\
		27 m     & 0.8297   & 0.9716    & 0.9071    & 0.9071     & 0.9062     \\
		53 m     & 0.9128   & 0.9854    & 0.9524    & 0.9524     & 0.9527     \\\hline  
	\end{tabular}
	\label{table:dist}
\end{table}
\vspace{-6pt}	
\begin{table}[b]
	\scriptsize
\vspace{-8pt}	\caption{Performance of orientation variation.}
	 \vspace{-5pt}
	\begin{tabular}{llllll}
		Angle\\(deg) & Accuracy & $F1_{micro}$ & $F1_{macro}$ & $AUC_{micro}$ & $AUC_{macro}$ \\ \hline
		0          & 0.989    & 0.9981    & 0.994     & 0.994      & 0.9939     \\
		30         & 0.9542   & 0.9923    & 0.975     & 0.975      & 0.975      \\
		60         & 0.9792   & 0.9965    & 0.9887    & 0.9887     & 0.9887     \\
		90         & 0.9819   & 0.9969    & 0.9901    & 0.9901     & 0.9902  
		\\ \hline  
	\end{tabular}
	\label{table:orient}
	\end{table}
	\begin{table}[b] \vspace{-5pt}
	\scriptsize
	\caption{Performance of MeRFFI though the wall.}
	\begin{tabular}{llllll}
		& Accuracy & F1\_micro & F1\_macro & AUC\_micro & AUC\_macro \\ \hline 
		Blockage & 0.9678  & 0.9946  & 0.9824  & 0.9824 & 0.9825 \\\hline  
	\end{tabular}
	\label{table:thruwall}
 \end{table}
 \begin{table}[b]
\vspace{-12pt}
	\scriptsize
	\caption{Classifying 24 signatures in 4 Channels (96 classes).\vspace{-5pt}}
	\begin{tabular}{llllll}
		Test-Train \\ Split & Accuracy & $F1_{Micro}$ & $F1_{Macro}$ & $AUC_{Macro}$ & $AUC_{Micro}$ \\\hline
		50\% test & 0.9429   & 0.9799   & 0.9865   & 0.97      & 0.96      \\
		60\% test & 0.9104   & 0.9866   & 0.9862   & 0.95      & 0.95      \\
		70\% test & 0.93     & 0.9875   & 0.9872   & 0.97      & 0.97      \\
		80\% test & 0.903    & 0.9746   & 0.9865   & 0.91      & 0.91      \\
		90\% test & 0.8454   & 0.9704   & 0.9862   & 0.92      & 0.92      \\\hline \newline
	\end{tabular}
	\label{96t}
\end{table}
\vspace{-7pt}
\subsection{Performance with Variation in Distance (Path Loss)}
\textbf{Objective:} This experiment is performed to test the variation in the performance of the MeRFFI system with varying distances. 

\textbf{Experimental setup:} This experiment is performed using the same COTS devices mentioned in the previous experiment.  Here the distance of the receiver is varied in each experiment (distance). The receiver is kept at three different distances 7 m, 27 m and 53 m from the receiver location and the experiment is done on WiFi's channel 5. The experimental setup is labeled as 'B' in the figure \ref{fig:davisplan2} and as seen in the figure, is a 53 m long corridor, thus a complex environment from the wireless channel perspective. Twelve different signatures were injected through the MeRFFI device on the transmitted wireless signal. Packets are transmitted for 20 seconds for each signature. The signatures received at the receiver are classified using a CNN whose structure is similar to the one shown in the table \ref{CNN1}. Here only 20 percent of the data is used for training and the remaining 80 percent is used for testing. 

\textbf{Result and inference:} Table \ref{table:dist} shows the performance of the MeRFFI system with different distances 7m, 27m, and 53m. It can be seen that the system performs well, despite the wireless channel being a complex tunnel-like environment. Hence MeRFFI's injected signature does not significantly vanish from the wireless physical layer with distance. Since MeRFFI's application scenario requires it to be kept very close to the transmitter, it can be inferred that the injected signatures exist at very short delay spread compared to other multipath components that come into picture when the communication distance increases.\vspace{-9pt}
\subsection{Impact of Orientation}
\textbf{Objective:}  This experiment is done to test whether a change in orientation between the transmitter and receiver affects the performance of MeRFFI's  RF-fingerprint injection capability. 

\textbf{Experimental setup: } This experiment was performed in the same COTS equipment mentioned in the previous experiment. The transmitting laptop has the Intel 8265ac is attached with the MeRFFI device and is kept at a fixed location. The receiver laptop is at a distance of 2.54 m from the transmitter. The experimental setup is marked 'D' in the figure  \ref{fig:davisplan2}. The position of the receiver is then varied by 30 degrees with the distance being the same. The same 12 voltage values are injected in four such orientations 0$^\circ$, 30$^\circ$, 60$^\circ$, and 90$^\circ$ respectively. This data is then classified using CNN for each location, and their performance is compared. Same as the previous experiment, 20 percent of the data is used for training, and the remaining 80 percent is used for testing.

\textbf{Result and inference:} Table \ref{table:orient} shows the performance of the MeRFFI system with varying orientation. The minimum accuracy obtained was 95.42 percentage. It can be seen that the change of orientation does not affect the performance of this system. \vspace{-13pt}	
\subsection{Through Wall Signature Injection}
\textbf{Objective:} This experiment is done to test the MeRFFI's signature injection capability when there is a blockage between the transmitter and the receiver.

\textbf{Experimental setup:} This experimental setup, which is represented as 'C' in the figure \ref{fig:davisplan2} is placed such that the transmitter is in one room, 100 m away from the wall and the receiver is placed 0.3 m away from the wall in the next room. There is no other line of sight available for these COTS devices to communicate. The same 12 codes as in the previous experiment are injected by the transmitter. The received codes are classified using the CNN classifier with the structure as given in table \ref{CNN1}. Similar to the previous cases, 20 percent of the data was used for training, and the remaining 80 percent was used for testing. 

\textbf{Result and inference:} As it can be seen from the table \ref{table:thruwall}, MeRFFI performs well through a wall. With 20 percent training and remaining test, MeRFFI has the potential to be applied for real-world scenarios with obstruction.\vspace{-8pt}
\subsection{Effect of Using Multiple Channels}
\textbf{Objective:} To test the hypothesis that the same signature injected by MeRFFI appears as a different one if the devices use a different channel for communication. 

\textbf{Experimental setup:} Here we use the test set up 'A' as shown in figure \ref{fig:davisplan2}. 24 signatures were injected in 4 different channels. Thus a 96 class recognition system had to be built.

\textbf{Result and inference} The minimum micro averaged AUC for test to train ratio varying from 50$\%$ to 80$\%$ was observed to be 0.95. Table \ref{96t} shows the averaged F1 scores and accuracy. From the observations, we can validate this hypothesis that MeRFFI's signature appears different when observed from a different channel number.
\vspace{-11pt}
\subsection{Throughput, Power, Cost and other Inferences}
All the above experiments were done using the very high throughput mode (VHT) of IEEE 802.11 ac. The data rate is always above 50 Mbps, thus indicating that the injected perturbations do not significantly hinder the datarate. We also did not observe any packet drops while  MeRFFI was attached to the transmitting node.

The cost of manufacturing our prototype was less than 10 dollars including the price of commercial varactor chips used. The cost was kept low due to in lab design and fabrication. On mass production these materials would cost less than 20 cents per device.

The entire power including the control mechanism, i.e., channel select and signature inject code was less than 50 millijoules (0.05 Ws), which, when compared to DSP chip is $67\%$ less power consumed. The energy consumption of a programmable MeRFFI module does not vary with packet size. In our COTS testbed, the above results indicate that MeRFFI is very suitable for commercial IoT adoption.\vspace{-6pt}

\section{Security Analysis}
In this section, we discuss whether our proposed system is vulnerable to a man-in-the-middle scenario and replay attack.\vspace{-12pt}
\subsection{Threat Model}
\begin{figure}[t]
\vspace{-5pt}
	\centering
	\includegraphics[width=0.8\columnwidth]{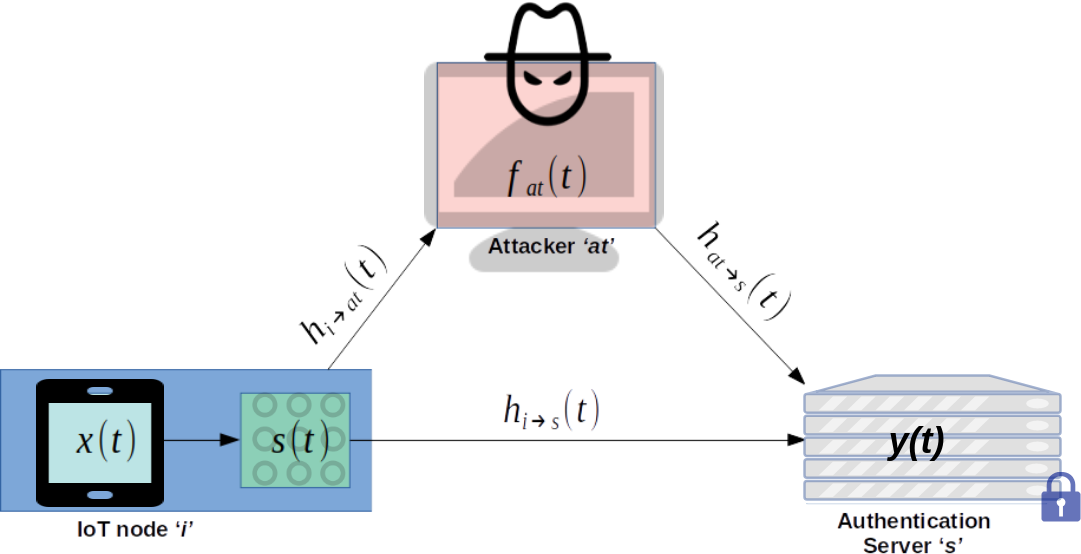}
	\vspace{-5pt}
	\caption{Communication between IoT node 'i' and server 's', where 'at' is trying to attack the system.}
 	\vspace{-14pt}
	\label{secal}                                          
\end{figure}

Consider an RF-Fingerprint authentication scenario, as seen in figure \ref{secal}, where an IoT node '$i$', which has the MeRFFI module attached, is trying to access an authentication server '$s$'. There is a man-in-the-middle scenario where an attacker '$at$' is trying to break this authentication mechanism. Here, '$at$' can try to impersonate the genuine node '$i$'.
Let $y_i(t)$ be the original signal from the node $i$ that is received at the server $s$. This can be represented as   
\begin{equation}\label{eq:re}
y_i(t)=x(t)*s(t)*h_{i\rightarrow s}(t)+\eta_{i\rightarrow s}
\end{equation}
where $s(t)$ is the transfer function of MeRFFI module and $h_{i\rightarrow s}(t)$ be the impulse response of the IoT node to the server channel. In comparison to this, $y_{at}(t)$ the received signal replayed by the attacker and received by the server can is given by:
\begin{equation}
y_{at}(t)=x(t)*s(t)*h_{i\rightarrow at}(t)*f_{at}(t)*h_{at\rightarrow s}(t)+\eta_{at}
\label{eq:at}
\end{equation}
Here $h_{i\rightarrow at}(t)$ is the impulse response of node $i$ to attacker $at$ channel and $h_{at\rightarrow s}(t)$ is the channel response from attacker to server. let $f_{at}$ be  the filter function of the attacker. From equations \ref{eq:re} and \ref{eq:at} we can see that in-order for the attacker to successfully replay the injected feature, the following condition should be satisfied:
\begin{equation}
s(t)*h_{i\rightarrow s}(t) = s(t)*h_{i\rightarrow at}(t)*f_{at}(t)*h_{at\rightarrow s}(t),
\end{equation}
\noindent
\textbf{Assumptions:} The attacker can launch the replay attack in mainly two ways \cite{Danev2010}, \textit{signal replay attack} and \textit{feature replay attack}.
In a signal replay attack, '$at$' records node $i$'s signal and simply re-transmits the signal without any modification. Here the attacker does not need to know the features that are injected. In a feature replay attack, the attacker does not know the injected features but is trying to derive such information, in which the filter function of the attacker $F_{at}$ is tuned such that the signal from the attacker $y_{at}(t)$ is same as that the original signal from node $i$, i.e., $y_{i}(t)$. \vspace{-8pt}
\subsection{Robustness Against Attacks}
\subsubsection{Against Signal Replay Attack :} From equation \eqref{eq:at} we can see that the attacker has no way of reproducing the physical channel by merely doing a signal replay attack, since $f_{at}(t)$, in this case, would be equal to a constant gain value $g$. The only case when a signal replay attack may work is when the attacker is very close (less than $\lambda/4$ distance from the tx or rx, also known as the guard zone \cite{He2016}). For 2.4 GHz WiFi, this would be 6.25 cm.

\noindent
\subsubsection{Against Feature Replay Attack :} Let the injected channel response $s(t)*h(t)$ be denoted as $sh$. Then the frequency domain representation of the filter function of the attacker can be derived from the previous equation as: 
\begin{equation}
F_{at}(a \exp ^{j\omega})= \frac{SH_{i\rightarrow s}(a_1 e ^{j\omega})}{SH_{i\rightarrow at}(a_2 e ^{j\omega})*SH_{at\rightarrow s}(a_3 e^{j\omega}) },
\label{eq:filter}
\end{equation}
This equation shows the difficulty of the attacker as it has to simultaneously and accurately model three different channel models and then use it to precisely find a filter response that when transmitted through the air would physically be equal to the injected signature by MeRFFI in node $i$'s signal. Looking this difficulty from a computational aspect: if we are using a 64 subcarrier WiFi signal, the attacker has to accurately model three channel responses. For each of the three channel responses, the attacker has to find a vector of 64 complex numbers, then use this vector to accurately model the injected signal response, which is another complex vector of size 64.\vspace{-5pt}
\begin{figure}[t]
\vspace{-5pt}
	\centering
	\includegraphics[height=1.99in, width = 3.4in]{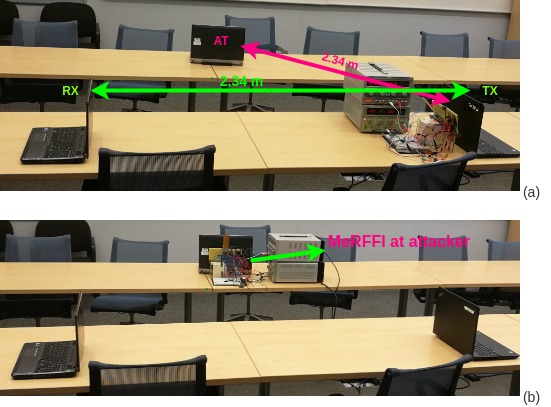}
	\vspace{-5pt}
	\caption{(a) Stage 1: Transmitter broadcasting packets using MeRFFI - attacker training its system (b) Stage 2: Attacker transmitting feature replay signatures with MeRFFI}
 	\vspace{-1pt}
	\label{fig:attack}                                          
\end{figure}
\begin{figure}[t]
\vspace{-10pt}
	\centering
	\includegraphics[width=0.88\columnwidth]{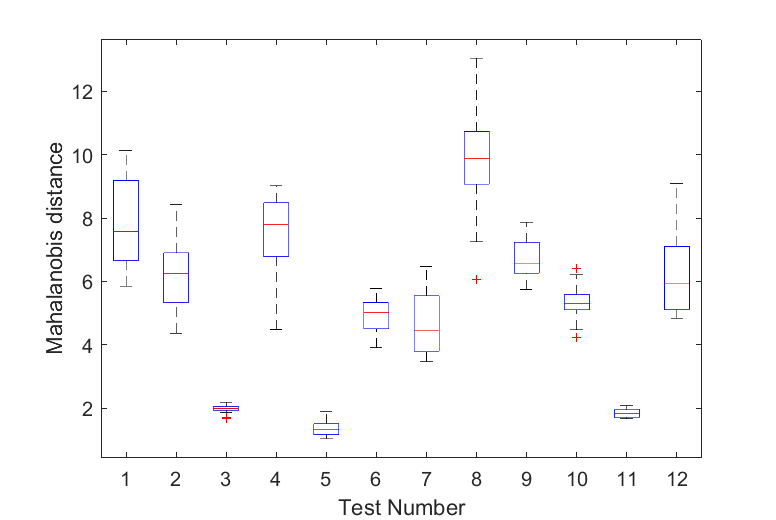}
	\vspace{-5pt}
	\caption{Figure showing the Mahalanobis distance between the CSI features of Alice and AT before classification, as seen by Bob. The Box plot depicts 12 different tests with 25 packets each} 
 	\vspace{-5pt}
	\label{fig:Maha}                                       
\end{figure}
\subsection{Experimental Validation:} 
\textbf{Objective:} To perform a feature replay attack on MeRFFI system.

\textbf{Experimental Setup:}
For this experiment there are three entities involved: \textit{(a)} the legitimate transmitter (Alice), \textit{(b)} the receiver (Bob) and \textit{(c)} the attacker (AT). Their specifications are as follows: \\
\textit{(a) Alice :} A Lenovo ThinkPad E570 equipped with an Intel 8265ac NIC card with MeRFFI attached as the transmitting node.\\
\textit{(b) Bob:} a Dell Latitude E6530 fitted with an Intel 5300ac NIC card as the receiver.\\
\textit{(c) AT} : In-order to give the attacker, the best possible capabilities, we equipped it with the best devices we have. We designed the AT with four devices, \textit{(i)} a Dell Latitude E6530 fitted with an Intel 5300ac NIC card, which acts as evesdropper for snooping the signal transmitted by Alice, \textit{(ii)} Dell Precision tower server-T5810 with Intel Xeon CPU E5-1630 @ 3.7 GHz and 64 GB ram, for performing the feature learning, similar to equation \ref{eq:filter} , \textit{(iii)} A Lenovo ThinkPad E570 equipped with an Intel 8265ac NIC card for doing the feature replay transmission \textit{(iv)} MeRFFI for injecting the features that were evesdropped.

The experimental setup can be seen in figure \ref{fig:attack}, and the distances between the entities are marked in the figure. In our setup, Alice transmits to Bob, AT is trying to perform a feature replay attack. AT has knowledge of the frequency channel used for communication, the channel control voltage $ CV$ to select that channel, and several signature control vectors $\vec{SV}$ (secret hardware key) and their responses. AT first receives the packet as an eavesdropper with equipment \textit{(i)}, then uses a deep learning-based supervised function approximator using the setup \textit{(ii)} to train estimate the value of  $\vec{SV}$ at the receiver, let the estimated value be  ($\vec{SV_r}$). AT then uses equipment \textit{(iii)} to transmit packets to Bob by using MeRFFI, which is controlled with the estimated $\vec{SV_r}$. Bob already has a classifier trained to accept only legitimate devices (here Alice's RFFs generated with MeRFFI). Bob receives the packets from Alice and the Attacker and checks them for authenticity.

To quantify the attacking scenario. First, Alice transmits 5500 packets of 11 different MeRFFI signatures. The Attacker knows all the 11 signature control vectors $\vec{SV_i}, i=1,2..11)$, uses these as labels to train its deep learning based function approximator (from the 5500 packets, 4950 packets are used to train the network and 550 are used to validate, this is to avoid overfitting). The function approximator is a modified version of CNN shown in table \ref{CNN1}, where the output layer is modified to get smooth estimates\cite{Ferrari05, ohn19}. With this we perform our evaluation with 2 test procedures.\vspace{5pt}
 
\textit{Test procedure 1:} We test the linear separability of the received data of 12 different attacks (transmissions by AT). The separability is measured by Mahalanobis distance between Alice's data and AT's data, as seen by Bob. In each scenario AT listens for 5500 packets from 11 $SV$ signals of Alice, and sends 25 attack signatures towards Bob.\vspace{5pt}

\textit{Test procedure 2:} In this test, we identify the CSI samples with the least distance from test procedure 1 and perform an attack on Bob. Bob has an authentication system built (using CNN shown in table \ref{CNN1} to accept all 12 signatures from Alice. Then Attacker uses its learned function approximator to generate 500 packets for which the signature control vector is $\vec{SV_x}$, where $x$ is a secret between Alice and Bob. This estimated control signal, $\hat{\vec{SV_x}}$  is used to send 500 packet burst with MeRFFI to force an authentication with Bob.

\textbf{Result:} From test procedure 1, we identified that test 5 has the least distance between Alice's signature and AT's signature, as seen by Bob. The distance, which is equal to 1, is still a good distance, which a strong classifier can classify. But this signature is the weakest among the one we tested. We then perform procedure 2, where we use this scenario to send 500 attack packets to Bob. Here Bob has an authentication system that is tailored to accept 12 different signatures from Alice. In this experiment, all the 500 packets from AT were rejected by Bob. We also checked if time variability affects the attack, hence we performed this attack again after one week, but the results were the same.

\textbf{Inference:} From this experiment, we can infer that even after knowing many of the signature control vectors $\vec{SV_x}$, it is extremely difficult to attack the MeRFFI system if the control code is not known. This opens up the avenue for developing a data-dependent device procedure where a digital signature and physical signature are dependent, further improving the security condition. Another inference we can make is from test procedure-1 (figure \ref{fig:Maha}). We can observe that certain attack signatures may have high similarity to the original injected CSI, as is the case with tests 5, 3, and 11. Although they were still rejected authentication by a carefully designed classifier at Bob, this weakness can be reduced using protocol 2 and 3, which uses multiple code injections to authenticate a node. Other than the attack scenario mentioned above, one may argue that the attacker may also have the capability to record the downlink channel without the signature, inverse it, and then compare it with the injected channel to isolate the signature. But reciprocity of the channel is usually an assumption to get approximate CSI, due to the detrimental changes in the transceiver components and the channel \cite{Senay09}, thus making this measurement inadequate for the attacker.
\section{Discussions}\label{discuss}

\subsection{Security-Reliability Trade-off}
The radio frequency fingerprint community has been trying to address the contradicting tradeoff between security and reliability \cite{Robyns2017}. With MeRFFI we have optimized the dimensions based on our theoretical deductions in section \ref{cdyna} to maintain an optimum between security and reliability based on the environment. \vspace{-7pt}
\subsection{Size Matters}
The metamaterial research community has been actively researching the \cite{Tang2014} area of Frequency Selective Surfaces (FSS). The Prototype created in our lab has been miniaturized to half the size of its original dimension. Size reduction up to $1/16$ of the original dimension is common in the metamaterial research community. This miniaturization is also accompanied by narrowed operating bandwidth of the frequency selective surface and reduced gain.  The narrow bandwidth is good for RFF injection, a smaller number of perturbations can be made in the operating frequency, but a  reduced gain can make the authentication system sensitive to environmental effects. Thus, while designing the size of MeRFFI, we have to take into consideration the application and the frequency of operation.\vspace{-7pt}

\subsection{Enabling Node Movement}
In MeRFFI's classifier design, we used a CNN that is robust in a static environment, but if the receiver has a change in orientation or is required to be moved from one place to another, it would have to be retrained in the system. This is because the feature representation may not separate (linear separability) the injected signature from the environmental multipath. This can be solved either by generalization using transfer learning approaches \cite{Wagle2012, Lu20}, or using 2 step transmission for authentication.  To further clarify, if the feature representation of the injected signatures made in one position/orientation is termed source domain data and the representation in another position/orientation is the target domain data, we would want the features learned from the source domain to be valid in the target domain. In statistical terms, we need to generalize the injected features learned by reducing the difference between the distributions of the source and target domain data. 

Another way to solve this problem is to use a two code transmission for authentication, where the difference in the injected CSI's \cite{Dunna20} at the receiver yields the actual code. Let the IoT node transmit three security signatures $S1$,$S2$, and $S3$. These signatures are influenced by channel H. At the receiver, let the received signature be $SH_1$, $SH_2$ and $SH_3$. Using $[SH_1-SH_2]$ and $[SH_1-SH_3]$ and using this difference to authenticate the TX node can make the system robust to movement. Our future work is focused on realizing this authentication scheme.\vspace{-7pt}
\subsection{Energy Harvesting FSS}
Our experimental results suggest that an active Frequency Selective Surface (FSS) offers much more secure physical layer signatures than a passive one. Although the active MeRFFI prototype made in the lab consumes very minimal power, low power IoT applications would need energy harvesting based FSS \cite{fssenergy2} that does not need an external battery. 
\section{Conclusions}
In this paper, we design and implement the MeRFFI, a small and inexpensive metasurface, which injects RF-fingerprints into the wireless physical layer. Through experiments on a COTS testbed, we have shown that these signatures are very reliable and have a large design space. MeRFFI consumes only very little power to produce distinguishable RFFs, thus making it very suitable for IoT applications. We have suggested three easily integrable protocols using MeRFFI, where we have suggested the scenarios where MeRFFI can be used in a passive form or active form depending on the priorities and needs of a specific IoT network. Our security analysis discusses how significantly more complex this system is to be cracked compared to the current state-of-the-art.\vspace{-3pt}

\bibliography{bibi2}{}
\bibliographystyle{IEEEtran} 

\begin{IEEEbiography}[{\includegraphics[width=1.1in,height=1.3in,clip,keepaspectratio]{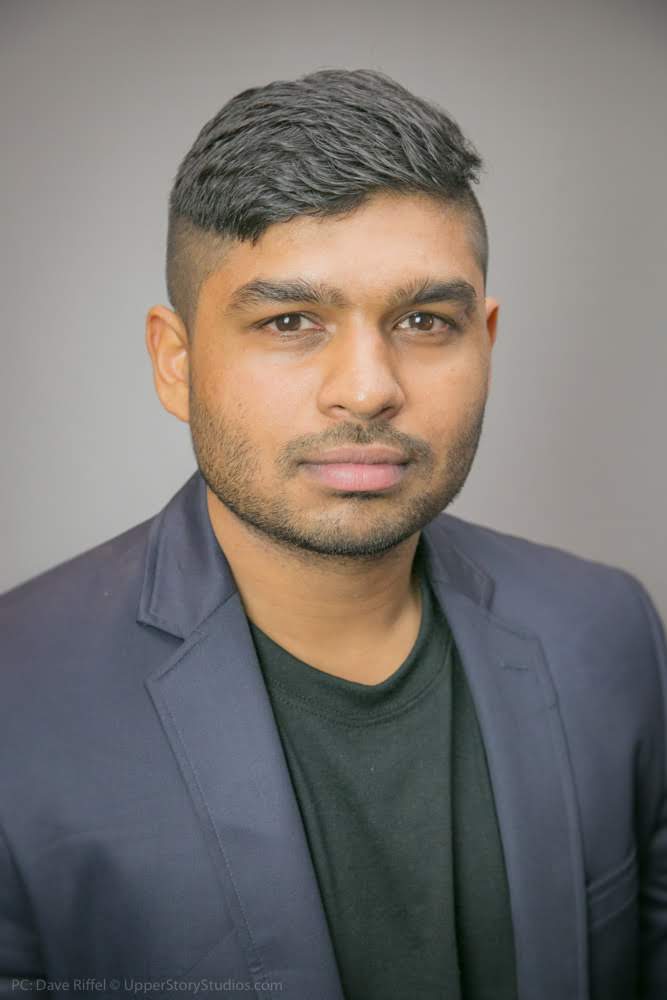}}]{Sekhar Rajendran (SM'19)} received his B.Tech. degree in electronics and communication from Cochin University of Science and Technology, India and his M.Tech. degree in signal processing and control systems from the National Institute of Technology, Hamirpur, India. Currently, he is a Ph.D. student in the Department of Electrical Engineering at the University at Buffalo, The State University of New York. Prior to this role, he was an Assistant Professor in the Electronics and Communication Department in Rajagiri School of Engineering and technology (India) and has also worked in the wireless industry as a Network subsystem Engineer for Sasken Technologies Limited. His current research interests are in wireless physical layer security, reconfigurable wireless communications and wireless sensing. He is a Student Member of the IEEE.
\end{IEEEbiography}

\begin{IEEEbiography}[{\includegraphics[width=1.1in,height=1.3in,clip,keepaspectratio]{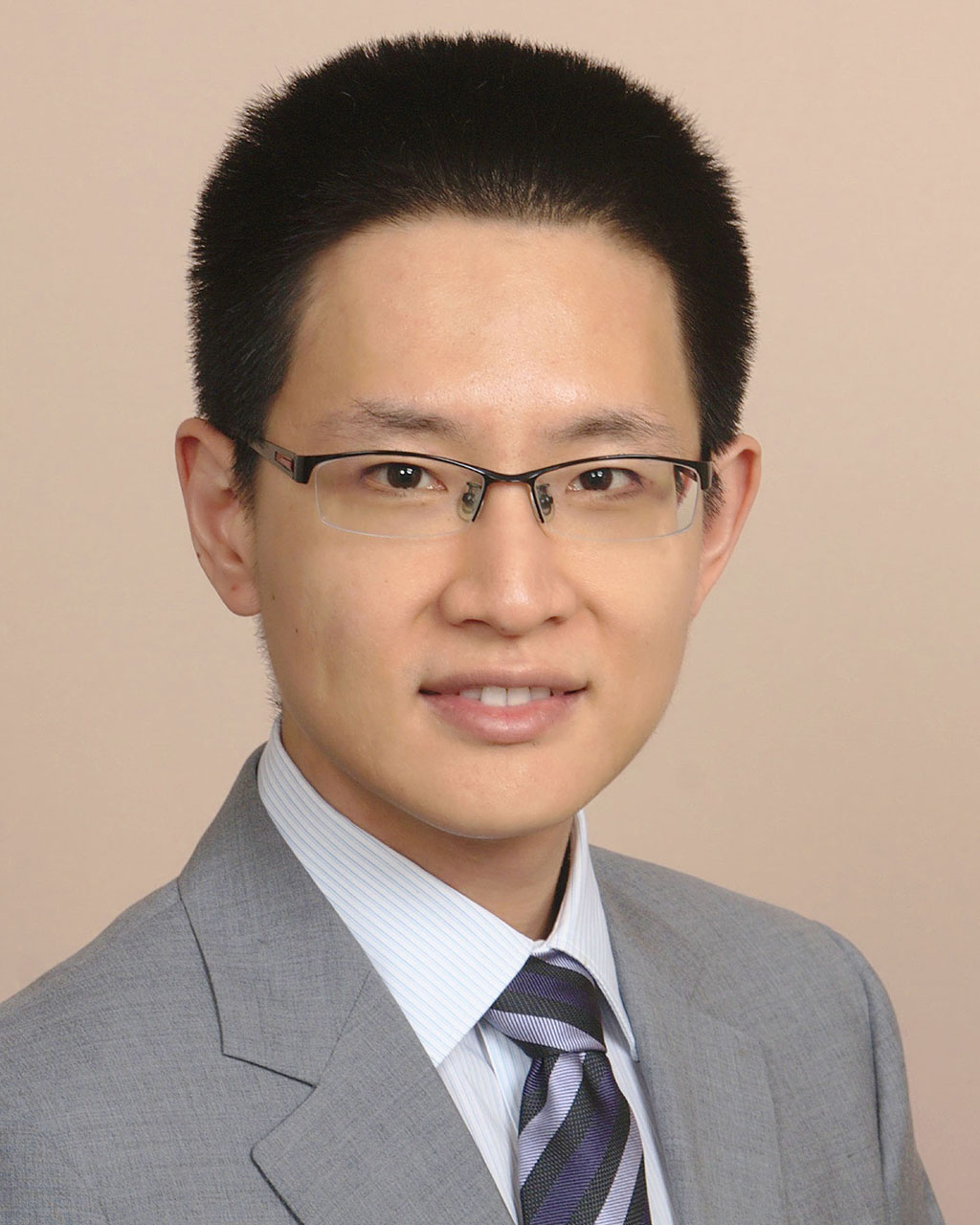}}]{Zhi Sun (S’06-M’11-SM’19)} received the B.S. degree in telecommunication engineering from Beijing University of Posts and Telecommunications (BUPT), Beijing, China, in 2004, the M.S. degree in electronic engineering from Tsinghua University, Beijing, in 2007, and the Ph.D. degree in electrical and computer engineering from the Georgia Institute of Technology, Atlanta, GA, USA, in 2011. He was a Postdoctoral Fellow with Georgia Institute of Technology from 2011 to 2012. In 2012, he joined the Department of Electrical Engineering, University at Buffalo, The State University of New York, Buffalo, NY, USA, as an Assistant Professor. He is currently an Associate Professor with University at Buffalo. His research interests include physical-layer security, wireless communication and networking in extreme environments, metamaterial enhanced communication and security, wireless intrabody networks, wireless underground networks, wireless underwater networks, and cyber-physical systems. He was a recipient of the NSF CAREER Award in 2017, the UB Exceptional Scholar—Young Investigator Award in 2017, the Best Demo Award at IEEE Infocom 2017, the Best Paper Award at IEEE Globecom in 2010, the BWN Researcher of the Year Award at the Georgia Institute of Technology in 2009, and the Outstanding Graduate Award at Tsinghua University in 2007. He serves as an Editor for the IEEE TRANSACTIONS ON WIRELESS COMMUNICATIONS and Computer Networks (Elsevier). He is a senior member of IEEE.
\end{IEEEbiography}

\begin{IEEEbiography}[{\includegraphics[width=1.1in,height=1.3in,clip,keepaspectratio]{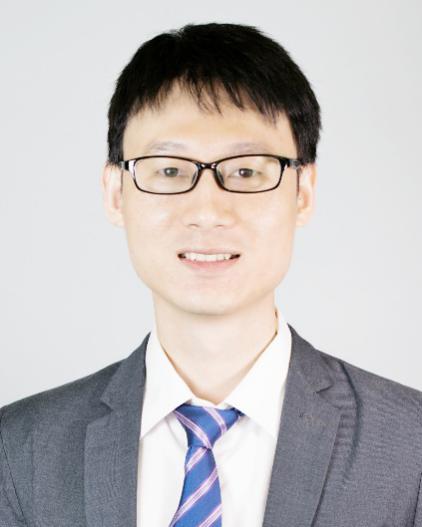}}]{Feng Lin (S’11-M’15-SM’20)} received the Ph.D. degree from the Department of Electrical and Computer Engineering, Tennessee Technological University, USA, in 2015. He is currently a Professor with the School of Cyber Science and Technology, College of Computer Science and Technology, Zhejiang University, China. He was an Assistant Professor with the University of Colorado Denver, USA, a Research Scientist with the State University of New York (SUNY) at Buffalo, USA, and an Engineer with Alcatel-Lucent (currently, Nokia). His current research interests include mobile sensing, Internet of Things security, biometrics, AI security, and IoT applications. Dr. Lin was a recipient of the Best Paper Awards from ACM MobiSys’20, IEEE Globecom’19, IEEE BHI’17, and the Best Demo Award from ACM HotMobile’18, and the First Prize Design Award from the 2016 International 3D printing competition.
\end{IEEEbiography}

\begin{IEEEbiography}[{\includegraphics[width=1.1in,height=1.3in,clip,keepaspectratio]{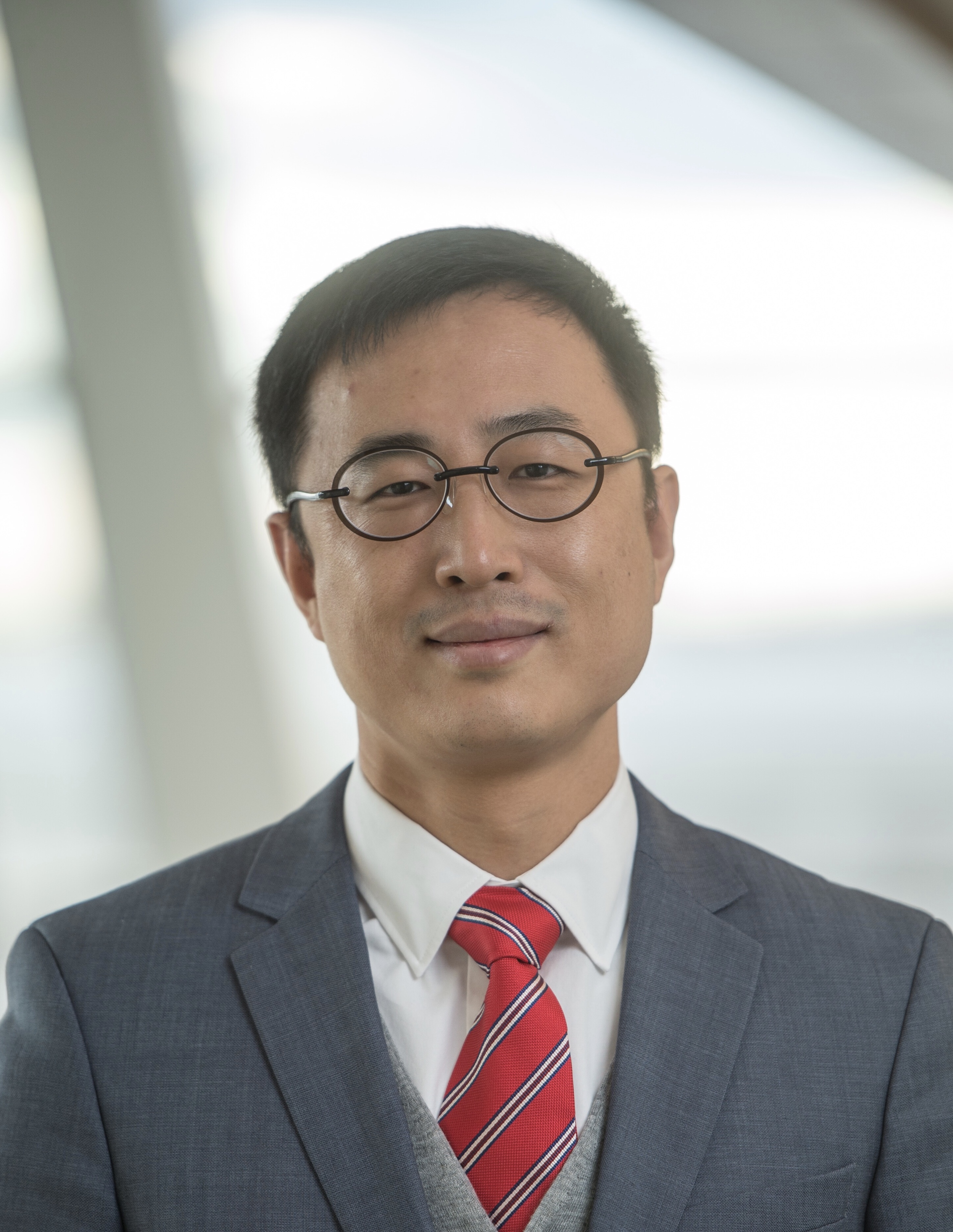}}]{Kui Ren (M’07–SM’11–F’16)} is Professor and Associate Dean of College of Computer Science and Technology  at Zhejiang University, where he also directs the Institute of Cyber Science and Technology.  Before that, he was SUNY Empire Innovation Professor at State University of New York at Buffalo. He received his PhD degree in Electrical and Computer Engineering from Worcester Polytechnic Institute. Kui’s current research interests include Data Security, IoT Security, AI Security, and Privacy. He received Guohua Distinguished Scholar Award from ZJU in 2020, IEEE CISTC Technical Recognition Award in 2017, SUNY Chancellor’s Research Excellence Award in 2017, Sigma Xi Research Excellence Award in 2012 and NSF CAREER Award in 2011. Kui has published extensively in peer-reviewed journals and conferences and received the Test-of-time Paper Award from IEEE INFOCOM and many Best Paper Awards from IEEE and ACM including MobiSys’20, Globecom’19, ASIACCS’18, ICDCS’17, etc. His h-index is 74, and his total publication citation exceeds 32,000 according to Google Scholar. Kui is a Fellow of IEEE, a Distinguished Member of ACM and a Clarivate Highly-Cited Researcher. He is a frequent reviewer for funding agencies internationally and serves on the editorial boards of many IEEE and ACM journals. He currently serves as Chair of SIGSAC of ACM China.

\end{IEEEbiography}

\end{document}